\documentclass{andp2012}% no class options needed by now
\usepackage[english]{babel}
\DeclareMathAlphabet{\mathcalnew}{OMS}{cmsy}{m}{n}
\usepackage{amsthm,amssymb}   % for math
\usepackage{comment}
\usepackage{graphicx,psfrag,subfigure}

\newtheorem{theorem}{Theorem}

\newcommand{\grafe}[1]{\left\{ #1 \right\}}
\newcommand{\tonde}[1]{\left( #1 \right)}
\newcommand{\quadre}[1]{\left[ #1 \right]}

\title{Review: Local Integrals of Motion in Many-Body Localized systems}
\author[J. Z. Imbrie]{John Z. Imbrie\inst{1}}
\author[V. Ros]{Valentina Ros\inst{2,4,}\footnote{Corresponding author\quad E-mail:~\textsf{vros@sissa.it}}}
\author[A. Scardicchio]{Antonello Scardicchio\inst{3,4}}
\address[1]{Department of Mathematics, University of Virginia, Charlottesville, VA 22904-4137, USA}
\address[2]{SISSA- International School for Advanced Studies, via Bonomea 265, 34136 Trieste, Italy}
\address[3]{The Abdus Salam ICTP, Strada  Costiera  11,  34151  Trieste,  Italy}
\address[4]{INFN Sezione di Trieste, Via Valerio 2, 34127 Trieste, Italy}
\shortauthors{J. Z. Imbrie, V. Ros, A. Scardicchio}
\begin{abstract}
 We review the current (as of Fall 2016) status of the studies on the emergent integrability in many-body localized models. We start by explaining how the phenomenology of fully many-body localized systems can be recovered if one assumes the existence of a complete set of (quasi)local operators which commute with the Hamiltonian (local integrals of motion, or LIOMs). We describe the evolution of this idea from the initial conjecture, to the perturbative constructions, to the mathematical proof given for a disordered spin chain. We discuss the proposed numerical algorithms for the construction of LIOMs and the status of the debate on the existence and nature of such operators in systems with a many-body mobility edge, and in dimensions larger than one.
\end{abstract}
\shortabstract
\begin{document}
\maketitle
% \noindent

\section{Introduction}

With the aim of describing the energy transport in spin systems, Anderson formulated in~\cite{anderson1958absence} a model of a quantum random walker in a stochastic potential landscape, now known as the ``Anderson model''. He argued that when the randomness is sufficiently strong, the quantum random walker is localized by the quenched disorder, meaning that localized initial conditions do not decay, and diffusive transport is suppressed. His theoretical work laid the foundations for the theory of the quantum dynamics in a strongly-disordered environment, with implications that go far beyond the realm of solid state physics; the occurrence of localization challenges indeed the basic assumptions underlying the theory of equilibration and thermalization in isolated quantum many-body systems~\cite{huse2015review, Eisert2016Rev}.

Whether localization occurs even in presence of interactions between the constituent degrees of freedom is a question of theoretical interest and practical relevance, which motivated the search for the so called Many-Body Localized (MBL) phase~\cite{Fleishman1980interactions}. The stability of the localized phase to the addition of weak scattering has been addressed theoretically in \cite{gornyi2005interacting, Basko:2006hh}, by means of a perturbative treatment applied to Hamiltonians of interacting fermions in a disordered potential (similar arguments have been extended to bosonic Hamiltonians \cite{Shlyapnikov2010bosons}). Subsequently, a large body of numerical works has revealed the occurrence of localization in one dimensional systems of interacting fermions on a lattice and spin chains in random fields~\cite{oganesyan2007localization, vznidarivc2008many, Bauer:2013rw, de2013ergodicity, mondragon2015,pal2010MBL,Kjall:2014fj,Vosk:2013kq, alet2015,Reichman2014Absence, John2015TotalCorrelations}. Signatures of this phenomenon have been found in the structure of the highly-energetic many-body eigenstates, for values of the interactions that lie outside the  perturbative regime. More recently, there are claims of experimental observations of MBL in artificial quantum systems of cold atoms \cite{Bloch2015, choi2016exploring,schneider2015} or trapped ions \cite{monroe2015}. Since these experiments focus on the suppression of transport, it is still debated how well they can distinguish MBL from single-particle Anderson localization (a review of the experimental situation can be found in another article in this issue). 

From the theoretical analysis it emerges that MBL systems exhibit a strongly non-ergodic dynamics, characterized by the suppression of diffusive transport of the global conserved quantities (such as energy, spin or particle number) and by the slow, power-law relaxation of local observables towards stationary values that are highly dependent on the initial condition. Remarkably, even in presence of interactions and at finite energy density, disordered quantum systems fail to thermally equilibrate following their own dynamics and remain permanently out-of-equilibrium. As such, they open interesting possibilities for the storage of quantum information, that can be locally manipulated and retrieved~\cite{serbyn2014interferometric}, for the protection of topological order at finite temperature, or for the realization of long-range order and finite temperature phase transitions in $d=1$, that would be forbidden by the equilibrium statistical mechanics~\cite{pekker2013hilbert,huse2013localization,chandran2013many}.

As it is well known, the breaking of quantum ergodicity is not an exclusive feature of disordered systems, as it is realized in ``conventional'' integrable systems satisfying the Yang-Baxter relations \cite{jimbo1990yang}. For such systems, ergodicity breaking is understood in terms of an extensive set of non-trivial conservation laws. It is natural to expect that a similar ``integrable'' structure emerges also in the disordered case; such a structure needs however to account for the suppression of diffusive transport, which is a distinguishing property of disordered systems. Many-Body Localized systems may thus be considered as a peculiar class of integrable systems, characterized by extensively many conserved operators (Local Integrals of Motion, or ``LIOMs'' in this review) whose structure in space prevents not only ergodicity and thermalization, but also transport over macroscopic scales~\cite{serbyn2013local, OganesyanHuseIntegrals, Swingle:2013ys}. In the following, we review the main efforts made to substantiate this perspective. 

This Review is structured as follows: In Sec. \ref{sec:phenomenology}, we recall the main features of Many-Body Localized systems. In Sec.\ref{sec:IOM} we discuss how such features can be justified in terms of the ``emergent integrability''. In Secs.  \ref{sec:analyticalConstruction} and \ref{sec:numericalConstruction} we review the main analytical and numerical constructions of conserved quantities that have been proposed in the literature. We devote Sec.\ref{sec:Discussion} to the discussion of debated issues concerning the fate of the conserved quantities in the case in which a mobility edge is present, and in dimensions higher than one.

\section{MBL systems: a multifaceted phenomenology}\label{sec:phenomenology}
In its most direct formulation, localization corresponds to the fact that local excitations do not decay. In the single particle setting, this was first shown to occur~\cite{anderson1958absence} for the Anderson model on a disordered lattice $\Lambda$, with Hilbert space $\emph{l}^2(\Lambda)$ and Hamiltonian
\begin{equation}
\label{eq:Hamiltonian}
H_{\text{A}}=\sum_{x \in \Lambda}\epsilon_x c^\dag_x c_x- \gamma \sum_{\langle x,y \rangle } \left( c^\dag_x c_y+c^\dag_y c_x \right),
\end{equation}
where $\langle x,y \rangle$ are edges in $\Lambda$, $\gamma$ is the kinetic (``hopping'') amplitude, and $\epsilon_x$ are independent random variables uniformly distributed in $\quadre{-W/2, W/2}$, defining a stochastic process indexed by the sites $x \in \Lambda$. It was argued in~\cite{anderson1958absence} that the level width of a local excitation at a site $x \in \Lambda$, given by $\Gamma_x(z)= -\Im S_x(z)$ with $S_x(z)$ the local self-energy defined by $G_{x}(z)= \langle x | (z-H_{\text{A}})^{-1}| x \rangle= ({z- \epsilon_x- S_{x}(z)})^{-1} $, goes to zero as the energy variable $z= E + i \eta$ approaches the real line. This holds in probability, i.e. for almost all realizations of the random landscape. It implies that the spectrum of local excitations remains discrete in the thermodynamic limit. 
An analogous statement can be formulated~\cite{Basko:2006hh, gornyi2005interacting} for many-body fermionic Hamiltonians of the generic form
\begin{equation}\label{eq:FermHam}
 H_{\text{MB}}= \sum_{\alpha} E_\alpha n_\alpha + \frac{1}{2} \sum_{\alpha \beta, \gamma \delta } \, U_{\alpha \beta, \gamma \delta} c^\dag_\alpha c^\dag_\beta c_\gamma c_\delta=H_0 + U,
\end{equation}
with $\alpha$ labeling the eigenstates of a quadratic part~\eqref{eq:Hamiltonian}. In this case, $\Gamma_a(z)$ is replaced with the imaginary part $\Gamma_{\alpha}(\epsilon, t)$ of the (Wigner transform of the) many-body self energy associated to the retarded Green function $G^R_{ \alpha} \tonde{t_1=t-\frac{\tau}{2}, \, t_2= t+ \frac{\tau}{2} }= -i \theta (t_2-t_1) \left\langle \grafe{c_{ \alpha}\tonde{t_1}, c^\dag_{\alpha}\tonde{t_2}} \right\rangle$, and the limit $\eta \to 0$ is replaced by the limit of vanishing coupling with an external thermal reservoir.
% \begin{equation}\label{eq:se}
%S^R_{\alpha}(\epsilon, t)= \int d\tau  e^{i \epsilon \tau}\, 
% S^R_{\alpha} \tonde{t- \frac{\tau}{2}, t+ \frac{\tau}{2}}
%\end{equation}
 % \begin{equation}\label{eq:GreenBAA}
%  G^R_{ \alpha} \tonde{t_1=t- \frac{\tau}{2}, \, t_2= t+ \frac{\tau}{2} }= -i \theta (t_2-t_1) \left\langle \grafe{c_{ \alpha}\tonde{t_1}, c^\dag_{\alpha}\tonde{t_2}} \right\rangle,
% \end{equation}. 
A vanishing typical value of $\Gamma_\alpha(\epsilon, t)$ implies that the irreversible evolution towards thermal equilibrium is hindered, as it follows from the fact that $\Gamma_\alpha(\epsilon, t)$ enters in the collision integral of the quantum Boltzmann equation: in absence of an external reservoir, the system fails to act as an heat bath for itself.

In the theoretical works~\cite{anderson1958absence,Basko:2006hh, gornyi2005interacting}, the above criteria given in terms of the level width of local excitations are recast as a problem of convergence (in probability) of the perturbative expansion for the decay rates around the trivially localized limits $\gamma=0$ in \eqref{eq:Hamiltonian} and $U=0$ in \eqref{eq:FermHam}. The mechanism for localization extends nonetheless also beyond the perturbative regime, as it has been shown by means of  numerical analysis of one-dimensional fermionic or spin Hamiltonians on the lattice. Prototypical models for the numerics are XXZ-chains in random longitudinal field \cite{pal2010MBL, alet2015, canovi2011quantum,de2013ergodicity}:
\begin{equation}\label{eq:XXZ}
\begin{split}
 H_{\text{XXZ}}&= \sum_{i =-K}^{K'-1} \quadre{J \tonde{S^x_i S^x_{i+1}+S^y_i S^y_{i+1}} + J_z S^z_i S^z_{i+1}}+ \sum_{i=-K}^{K'} h_i S^z_i, 
 %\\
 %&\equiv\sum_{i \in \Lambda}H_i,
 \end{split}
\end{equation}
or Ising chains in random transverse \cite{Kjall:2014fj} and longitudinal fields, among which the model considered in \cite{imbrie2014many}:
\begin{equation}\label{eq:HamImbrie}
H_{\text{I}} = \sum_{i=-K}^{K'} h_i S_i^z + \sum\limits_{i=-K}^{K'} \gamma_i S_i^x + \sum\limits_{i=-K}^{K'-1} J_i S_i^z S^z_{i + 1}.
\end{equation}
In both cases, $S_i^{\alpha}$, $\alpha= x,y,z$ are Pauli matrices on sites of a lattice $\Lambda = [-K,K'] \cap \mathbb{Z}$, with $S_i^\alpha \equiv 1$ for $i \notin \Lambda$, $\gamma_i = \gamma\Gamma_i$ with $\gamma$ small and the random variables $h_i, \Gamma_i, J_i$ are independent and bounded, with bounded probability densities.

 As a result of the effort to characterize MBL for these models, several diagnostics have been developed and exploited. In the following, we briefly recap the main features that are commonly considered as a signature of this phase; as we shall argue in Sec. \ref{sec:IOM}, all of these features can be justified in terms of the ``emergent integrability'' of the localized system, which thus furnishes a comprising definition of MBL.

{\bfseries \emph{(i) Absence of DC transport }} In the single particle case, Anderson's criterion on the level width of local excitations implies that local states $|x\rangle \in \emph{l}^2(\Lambda)$ are bound states. Stronger statements on the dynamics of localized initial conditions $|x\rangle$ can be rigorously proved, referred to as ``dynamical localization''. They correspond to the following bound holding almost surely:
\begin {equation}\label{eq:dynLoc}
 \sup_{t \geq 0} \tonde{\sum_{y} |y|^{2q} |\langle y | e^{-i H t}| x \rangle|^2} < \infty,
\end {equation}
implying that localized initial conditions $|x\rangle$ have, with probability one, uniformly in time bounded moments of all orders $q>0$. Such a condition rules out the possibility of transport, in particular of diffusive transport (which is instead expected in the weak disorder regime for $d \geq 3$).
The proof of~\eqref{eq:dynLoc} makes use of exponential bounds on either the local matrix elements of the resolvent~\cite{frohlich1983}, or on their fractional moments~\cite{aizenman1993}, or on the eigenstates correlators~\cite{Imbrie2016b}. Such bounds encode the exponentially-decaying structure of the eigenstates $\phi_\alpha$ of \eqref{eq:Hamiltonian}, whose envelope satisfies
\begin{equation}\label{eq:ExpDec}
 |\phi_\alpha(x)| \sim A_\alpha \, \text{exp}\tonde{- \frac{|x-r_\alpha|}{\xi_\alpha}}
\end{equation}
 where $r_\alpha$ is the localization center of $\phi_\alpha$ and $\xi_\alpha$ its localization length.

 Arguments for the vanishing of the diffusion constant are given also for the many-body case~\cite{Basko:2006hh, gornyi2005interacting}; similarly to the single particle case, they rely on the exponential decay of the correlations of the local density operators $\rho_r=c^\dag_r c_r$ on the exact many-body eigenstates $|E_n\rangle$:
\begin{equation}\label{eq:ExpCorr}
 \mathcalnew{L}_{n m}^{\rho}(r) = \sum_{r'} \langle E_n| \rho_{r'}| E_m\rangle \langle E_m| \rho_{r'+r}| E_n \rangle \lesssim \text{exp}\tonde{-\frac{|r|}{\xi(E)}},
\end{equation}
where $E_n \approx E_m \approx E$ and $\xi(E)$ is an energy dependent localization length. The absence of diffusion has been investigated numerically by analyzing the low-frequency behavior of the dynamical conductivity and its dc limit \cite{barivsic2016dynamical,berkelbach2010conductivity} and the spin-spin or density-density correlation functions in the infinite-time limit \cite{pal2010MBL,lev2014dynamics,vznidarivc2008many}. The vanishing of the transport coefficients is a peculiar feature of MBL systems, which contrasts with the efficient transport properties of clean integrable systems \cite{Mazur1969Non, SUZUKI1971ergodicity, Zotos1997transport}. Nevertheless, it is not an exclusive feature of this phase, as in one dimension the delocalized phase also displays sub-diffusive transport~\cite{reichman2014dynamics, AgarwalAnomalousDiffusion, Znidaric2016Diffusive, Vipin2015, Khait2016SpinTransport}.

{\bfseries \emph{(ii) Anderson localization in configuration space. }}The perturbative treatment in~\cite{Basko:2006hh, gornyi2005interacting} is consistent with the picture, originally proposed in the influential work~\cite{altshuler1997quasiparticle} (see also the construction in Sec. \ref{sec:PerturbationTheory} in this review), %and exploited in \VR{add here}\cite{Bauer:2013rw, Ponte2015196},
of MBL as Anderson localization on an abstract graph whose sites correspond to the Fock states diagonalizing the non-interacting part of \eqref{eq:FermHam}, and whose geometry is determined by the interactions. A similar picture can be formulated for the spin Hamiltonians \eqref{eq:XXZ} and \eqref{eq:HamImbrie}, with the Fock states replaced by the tensor products of the simultaneous eigenstates of the operators $S_i^z$. 
This scenario entails that the many-body eigenstates are ``weak deformations'' of the non-interacting states, a point of view emphasized in \cite{Bauer:2013rw}. This picture is corroborated by several numerical diagnostics, such as the scaling of the IPR of MBL eigenstates in the basis of non interacting  states~\cite{de2013ergodicity, alet2015}, or the expectation values of the non-interacting occupation numbers or $S^z_i$ operators, which are shown to be close to their non-interacting value also on MBL eigenstates~\cite{pal2010MBL}. The latter statement is rigorously proven in \cite{imbrie2014many} for the chain \eqref{eq:HamImbrie}, see Theorem \ref{thm:1} in Sec. \ref{sec:imbrie}.

{\bfseries \emph{(iii) Area-law entanglement in highly excited states. }} A key implication of the above scenario is the low entanglement entropy in the excited eigenstates $|E_n \rangle$ of MBL Hamiltonians. For one-dimensional systems, this is captured by the bipartite eigenstate entanglement entropy, obtained by splitting the chain into a right half R and a left half L, and by tracing the degrees of freedom corresponding to one of the halves,
\begin{equation}\label{eq:bipartite1}
 S= - \text{Tr} \tonde{\rho_R \log_2 \rho_R} 
\end{equation}
with 
\begin{equation}\label{eq:bipartite2}
 \rho_R= \text{Tr}_{L} \tonde{| E_n \rangle \langle E_n|}.
\end{equation}

    \begin{figure*}[!htb]
    \centering
       {\includegraphics[scale=0.5]{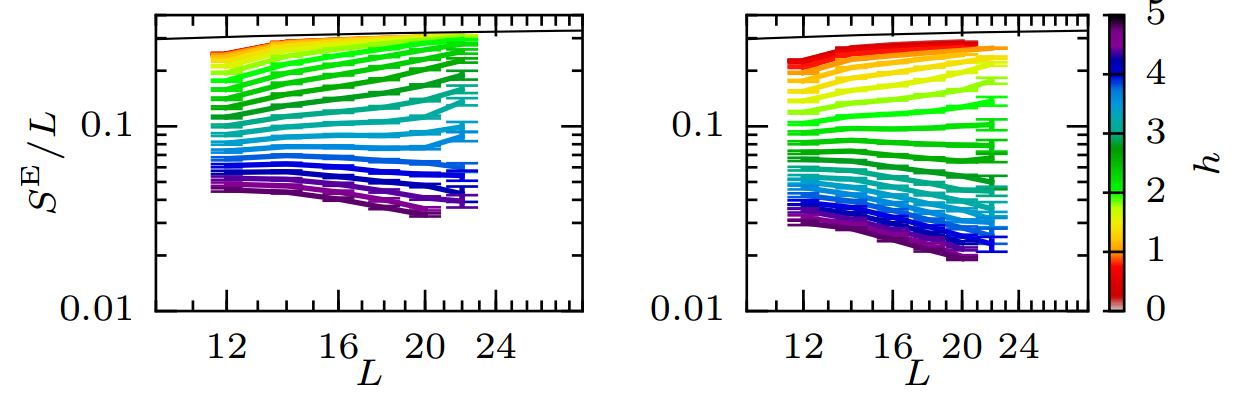}} 
        \caption{Bipartite entanglement entropy $S^E$ defined in Eq. \eqref{eq:bipartite1} for the Hamiltonian \eqref{eq:XXZ} with $J=J_z=1$ and $h_a \in \quadre{-h, h}$, as a function of system size $L$ for different disorder strengths in the middle of the spectrum (left) and in the upper part (right). For strong disorder, $S^E/L$ decreases signaling area-law. The figure is taken from Ref.~\cite{alet2015}}
  \label{fig:Bipartite}
    \end{figure*}

It is shown in~\cite{Bauer:2013rw, alet2015, Kjall:2014fj} that $S$ does not scale with system size but obeys an area-law, see Fig. \ref{fig:Bipartite}.
This is a typical ground-state property for generic gapped Hamiltonian, which in the MBL case extends to the whole spectrum. It implies that MBL eigenstates of extensive energy can be efficiently represented via Density-Matrix-RG or Matrix Product States \cite{YuMPS2015,Friesdorf2015,Khemani2015obtaining, Serbyn2016PowerLaw} and Tensor Networks \cite{Chandran2015spectral}. Moreover, it allows to access to the dynamical properties in the MBL phase adapting computational approaches originally designed for the ground-state physics, such as the strong disorder renormalization group exploited in~\cite{pekker2013hilbert} to 
construct approximate many-body eigenstates, and in \cite{Vosk:2013kq,vosk2013dynamical} to describe the dynamical evolution of initial product states. 
Further signatures of MBL are encoded in the structure of the full entanglement spectrum \cite{Yang2015TwoComponents,Geraedts2016entspectrum, Serbyn2016PowerLaw}: in particular, a distinctive feature of MBL is the power-law decrease  \cite{Serbyn2016PowerLaw} of the size of the eigenvalues of the reduced density matrix \eqref{eq:bipartite2}.
%, which was developed in~\cite{DasguptaMa1980,Fisher1992, MonthusIgloi} to capture the low-temperature thermodynamical properties of random magnets, and has been recently exploited to characterize the MBL phase by either constructing approximate many-body eigenstates~\cite{pekker2013hilbert} or by describing the dynamical evolution of initial product states~\cite{Vosk:2013kq,vosk2013dynamical}.  

{\bfseries \emph{(iv) Violation of the Eigenstate Thermalization Hypothesis. }} 
The structure of the MBL eigenstates is inherently incompatible with the Eigenstate Thermalization Hypothesis (ETH), formulated in~\cite{JensenETH,Deutsch1991, Srednicki1994,Rigol2008} as a microscopic justification for quantum thermalization. The ETH conjectures that the individual eigenstates $|E_n \rangle $ of a thermalizing Hamiltonian locally reproduce the thermodynamic ensembles, meaning that the eigenstates expectation values $\langle E_n | {\rm O} | E_n \rangle$ of local observables ${\rm O}$ depend smoothly on the energy of the state, and coincide with the microcanonical expectation values at energy $E\approx E_n$. This hypothesis, together with the assumption of exponential decay (with the system size) of the off-diagonal matrix elements $\langle E_n| {\rm O} | E_m \rangle$, guarantees that out-of-equilibrium initial state relax to states that are ``locally thermal''. 

\begin{figure*}[h!t]
   %\captionsetup[subfigure]{labelformat=empty}
  \centering
        \subfigure[]{\includegraphics[scale=0.28]{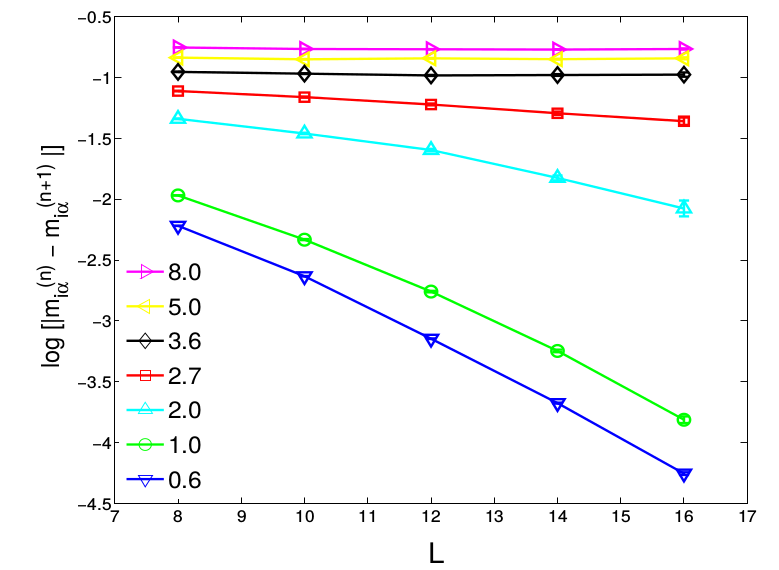}}   
        \subfigure[]{\includegraphics[scale=0.28]{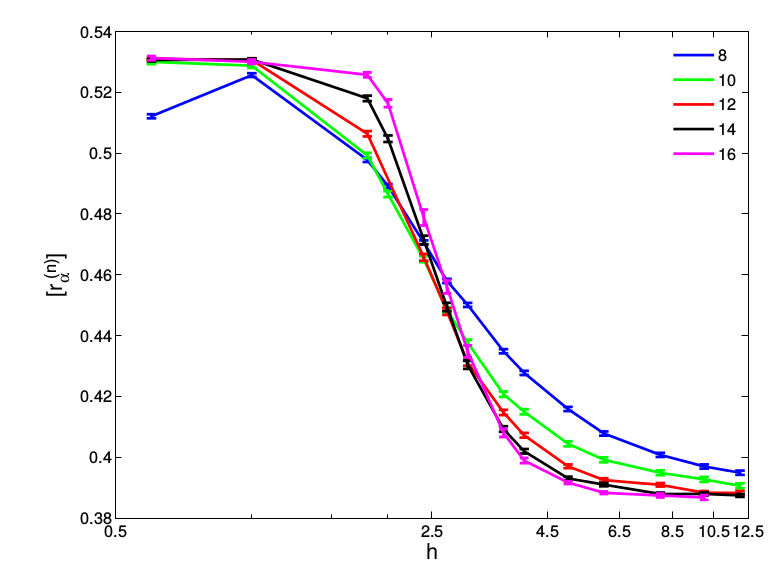}}
        \caption{(a) Logarithm of the averaged difference between the local magnetizations $m_{i \alpha}^{(n)}= \langle n | S^z_i | n \rangle_{\alpha}$ in adjacent eigenstates of the Hamiltonian \eqref{eq:XXZ} with $J_z=J=1$ and $h$ indicated in the
legend. The average is over the disorder realization $\alpha$ and the pairs of eigenstates. For large $h$, the differences remain large as the length of the chain L
is increased.  (b) Ratio of adjacent energy gaps defined in Eq. \eqref{eq:rparameter} for different system sizes $L$ indicated in the legend. For large $h$, the level statistics are Poisson. Figures taken from Ref.~\cite{pal2010MBL}}
  \label{fig:ETH}
    \end{figure*}

Contrary to the ETH requirements, MBL eigenstates in the same energy shell are locally distinguishable and non-thermal: the expectation values of the local observables are far from their equilibrium value, and they strongly fluctuate between states that are close in energy \cite{PeriodicallyPRL}, see Fig. \ref{fig:ETH}(a). ETH is also incompatible with the area-law scaling of the bipartite entanglement, as it requires the entanglement to be  equal to the thermal equilibrium entropy of the subsystem, that scales with the number of its degrees of freedom. This dichotomy has been exploited extensively to pinpoint the phase diagram of disordered systems by probing the violation of ETH in individual eigenstates obtained from the exact diagonalization; it is however useful to keep in mind that while localization is a statement on the long-time limit of thermodynamically large systems, the ETH ansatz determines the stationary behavior of finite-size samples: these two approaches are not necessarily equivalent \cite{Chandran2016beyond}.

{\bfseries \emph{(v) Absence of level repulsion. }} Starting from the earlier works~ \cite{Berkovitz1999LevelStatistics, Avishai2002LevelStatistics}, the absence of level repulsion, generically regarded as a signature of integrability~\cite{BerryTabor,Bohigas1984Char,Poilblanc1993}, has been extensively exploited as a characterization of the MBL phase, see also \cite{canovi2011quantum,cuevas2012level, Avishai2002LevelStatistics,pal2010MBL, Modak2014LevelStat, serbyn2016level}. 
An extensively used indicator of the absence of repulsion \cite{oganesyan2007localization} is the average value $\langle r \rangle$ of the dimensionless ratio 
\begin{equation}\label{eq:rparameter}
 r_n= \frac{\min \grafe{E_{n+1}- E_n,\; E_{n+2}- E_{n+1}}}{ \max \grafe{E_{n+1}- E_n,\; E_{n+2}- E_{n+1}}}
\end{equation}
obtained from the gaps between consecutive eigenvalues. In the localized phase, for increasing system size the latter approaches the Poisson theoretical value $\langle r \rangle \approx 0.39$, see Fig. \ref{fig:ETH}(b).

\begin{figure*}[!htb]
   %\captionsetup[subfigure]{labelformat=empty}
  \centering
        \subfigure[]{\includegraphics[scale=0.24]{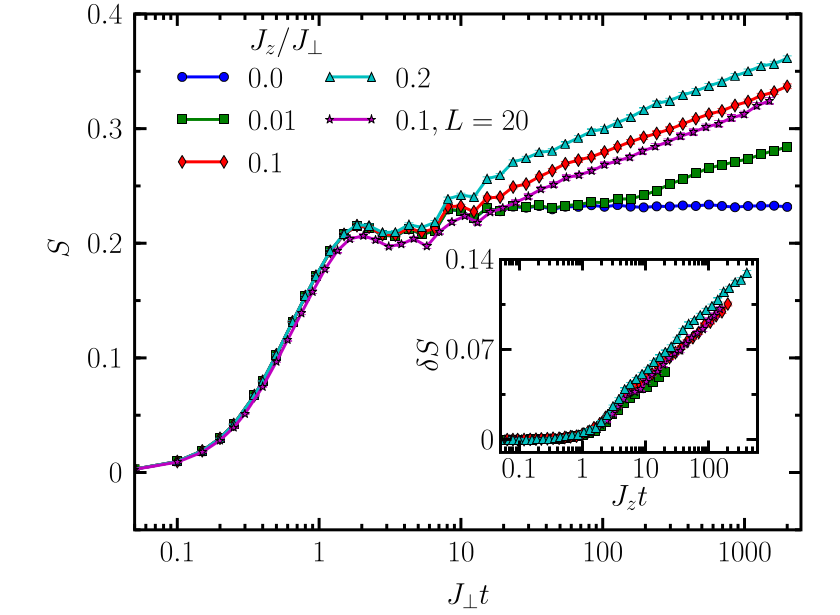}} \hspace{.3 cm}
        \subfigure[]{\includegraphics[scale=0.24]{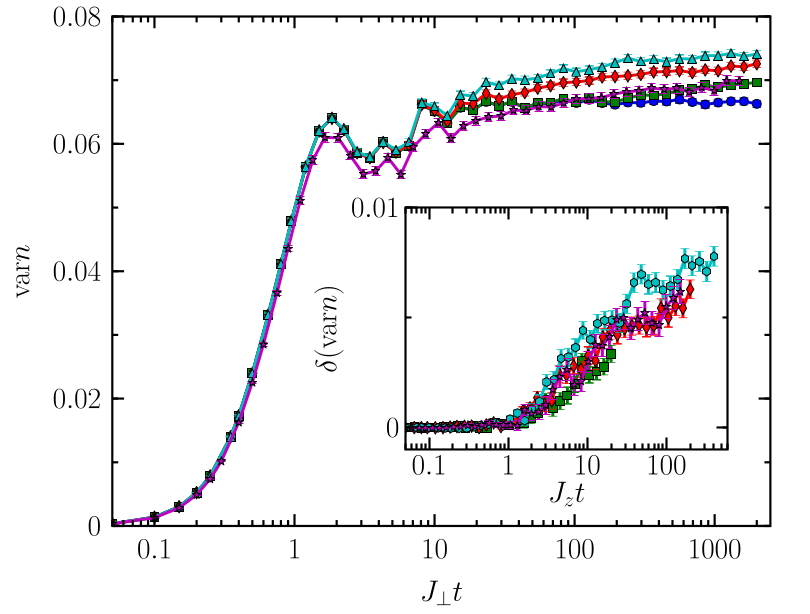}}
        \caption{(a) Unbounded growth of the bipartite entanglement after a quench starting from a site-factorized $S^z$ eigenstate of the Hamiltonian \eqref{eq:XXZ} with $J=J_\perp$, $h_a \in \quadre{-5, 5}$, $L=10$ and different interaction strengths $J_z$. The inset shows the same data with a rescaled time axis and subtracted $J_z = 0$ values. (b) Growth of the particle number fluctuations of a half chain after the quench. The behavior is qualitatively different than the entanglement entropy: the interactions do enhance the particle number fluctuations, but while there are signs of a  logarithmic growth as for the entanglement, this growth slows down with time. Figures taken from Ref.~\cite{bardarson2012unbounded}}
  \label{fig:bardarson}
    \end{figure*}

{\bfseries \emph{(vi) Slow growth of entanglement and slow relaxation. }}Despite the structure of MBL eigenstates is not significantly altered by the interactions, the latter strongly affect the quantum dynamics. Their signatures are traced in the real-time evolution of the bipartite entanglement entropy $S(t)$,  obtained as in~\eqref{eq:bipartite1}, with the substitution of $|E_n \rangle$ in~\eqref{eq:bipartite2} with the time-evolved pure state $|\psi(t) \rangle$ of the entire system.  Numerical simulations performed on disordered spin chains initialized in a low-entangled/product state~\cite{DeChiara2006, vznidarivc2008many, bardarson2012unbounded} display a transient fast growth of $S(t)$ (dominated by the direct nearest-neighbor interactions across the cut and extending to times of the order of the inverse interaction coupling), followed by a slow logarithmic growth, see Fig. \ref{fig:bardarson}(a). The increase in entropy is expected to continue indefinitely for an infinite system, while for finite systems $S(t)$ saturates to a value which depends on the initial state only, it scales with the system size but it is nevertheless smaller than the one to be expected in the thermal regime~\cite{nanduri2014entanglement, Serbyn:2013he}.  

The logarithmic scaling is a peculiar feature of the MBL phase \footnote{As opposed to the linear growth in clean integrable or non-integrable systems~\cite{CalabreseCardy2005, Kim2013Ballistic} and to the sub-ballistic growth in disordered, delocalized systems close to the MBL phase~\cite{Luitz2016Extended, AltmanTheory2015, Potter2015Universal}.}, that allows to distinguish it from the non-interacting localized phase, where $S(t)$ saturates to a finite value independent of the system size. The unbounded growth of entanglement is perfectly compatible with the absence of transport, see Fig. \ref{fig:bardarson}(b). It is ascribed to the interaction-induced dephasing between the eigenstates involved in the decomposition of the initial product state. The same mechanism is at the root of other distinguishing dynamical features of MBL systems, such as the power-law decay in time of the response to a (properly designed) spin-echo protocol~\cite{serbyn2014interferometric}, the power-law relaxation (towards non-thermal values) of the expectation value of local observables~\cite{serbyn2014quenches} or of the concurrence (a measure of the non-classicality of the system) \cite{iemini2016signatures} after a quench, and the suppression, with respect to the non-interacting case, of the revival rates of single-site observables~\cite{moore2014revivals}.\\
 
We discuss in the following section how these different aspects of MBL can be understood by assuming the existence of extensively-many, non-trivial conservation laws. The discussion is done assuming the \emph{full} many-body spectrum is localized, irrespectively of the energy density. This assumption applies to lattice Hamiltonians having a bounded local spectrum, as pointed out in \cite{oganesyan2007localization}; the corresponding systems are often called```fully MBL'' in the literature. Comments on the occurrence of mobility edges are postponed to Sec. \ref{sec:Discussion}.

\section{Quasilocal integrals of motions: a unifying framework}\label{sec:IOM}
It is a common expectation that the failure of ergodicity in closed, interacting systems is related to some form of integrability. Indeed, the phenomenology of MBL systems summarized in Sec.\ref{sec:phenomenology} suggests that conservation laws exist for strongly disordered systems. This expectation has been made concrete in~\cite{serbyn2013local, OganesyanHuseIntegrals, Swingle:2013ys}, where it was suggested that MBL Hamiltonians are non-linear functionals of a complete set of LIOMs $I_\alpha$, of the form
\begin{equation}\label{eq:EffMod}
 H_{diag}= h_0 + \sum_{\alpha} h_\alpha \, I_\alpha + \sum_{\alpha, \beta} h_{\alpha  \beta}\; I_\alpha I_\beta+ \sum_{\alpha, \beta, \gamma} h_{\alpha  \beta \gamma}\; I_\alpha I_\beta I_\gamma + \cdots,
\end{equation}
where the dots stand for higher order products. The $I_\alpha$ are expected to be functionally independent and mutually commuting. The set is complete in the sense that every many-body eigenstate can be labeled in a unique way with the eigenvalues of the $I_\alpha$.

While the expansion~\eqref{eq:EffMod} is to some extent generic (in particular, to specify $H_{diag}$  it is necessary to determine a number of coefficients which scales with the size of the Hilbert space \cite{monthus2015integrals}), the fingerprint of localization is the ``\emph{quasilocality}'' of the operators $I_\alpha$. The notion of quasilocality extends, at the operator level, the structure of the single-particle eigenstates $\psi_\alpha$ of~\eqref{eq:Hamiltonian}, cfr. Eq. \eqref{eq:ExpDec}: the operator norm of $I_\alpha$ is expected to decay exponentially away from a compact region of space of a typical size $\xi_{op}$ around a center $R_\alpha$. More precisely, let 
\begin{equation}\label{eq:OpExp}
 I_\alpha= \sum_{\mathcalnew{I}} \mathcalnew{A}^{(\alpha)}_\mathcalnew{I} \, O_\mathcalnew{I} 
\end{equation}
be the expansion of $I_\alpha$ on a basis of local operators $O_\mathcalnew{I}$ labeled by $\mathcalnew{I}$. For the spin chains \eqref{eq:XXZ} and \eqref{eq:HamImbrie}, a suitable basis is made by the tensor products of local spin operators $S^{\alpha_1}_{i_1} \otimes \cdots \otimes S^{\alpha_n}_{i_n}$ with $\alpha_i \in \grafe{x,y,z}$ and $i_k \in \Lambda$, while in the fermionic case \eqref{eq:FermHam} the normal-ordered tensor products of creation and annihilation operators of single-particle states can be considered, see Eq. \eqref{eq:normOrd}. Let $S(\mathcalnew{I})$ denote the ``support'' of $O_\mathcalnew{I}$, i.e. the set of sites/local degrees of freedom on which the operators acts non-trivially. Quasilocality entails that
\begin{equation}\label{eq:QuasiLoc}
 |\mathcalnew{A}^{(\alpha)}_\mathcalnew{I}| \lesssim C_{\alpha}\, \text{exp} \tonde{- \frac{d\quadre{R_\alpha, S(\mathcalnew{I})}} {\xi_{op}}},
\end{equation}
where $d\quadre{R_\alpha, S(\mathcalnew{I})}$ is the distance between $R_\alpha$ and the furthest degree of freedom in $S(\mathcalnew{I})$. This encodes the property of the $I_\alpha$ of being weak deformations of the local, physical degrees of freedom (such as the local density or spin operators).

This spatial structure implies that the expansion~\eqref{eq:EffMod} is not structureless, as the typical value of the coefficients $h_{\alpha \beta \cdots}$ decays exponentially in the distance between the localization centers of the corresponding operators, on a typical scale that we denote with $\xi_{int}$. This structure is at the root of the genuinely many-body dynamical features of MBL systems, as it entails that the dephasing between the $I_\alpha$ induced by the interactions in \eqref{eq:EffMod} occurs over a broad range of time scales.

The phenomenological models in~\cite{serbyn2013local, OganesyanHuseIntegrals, Swingle:2013ys}  additionally assume that (i) the $I_\alpha$ have binary spectrum, and thus can be considered as effective spins or occupation number operators (also termed ``logical bits'' or ``l-bits''), and (ii) the full spin (or fermionic) algebra can be constructed, with ladder operators $I_\alpha^{\pm}$. This follows straightforwardly from the construction in~\cite{imbrie2014many}, where the conserved quantities of the disordered chain \eqref{eq:HamImbrie} are obtained from the Pauli operators $S_a^z$ by means of a quasilocal rotation $\Omega$,  $I_\alpha= \Omega S_a^z \Omega^*$ and $I^{\pm}_\alpha= \Omega S_a^\pm \Omega^*$, that obviously preserves the spectrum and the commutation relations. 

This set of assumptions is sufficient to derive the full phenomenology of MBL systems, as we shortly recap in the following.

\begin{itemize}
 \item[(i)] \emph{Absence of DC transport: } we report the argument given in \cite{rms-IOM}. The Kubo formula for the dc conductivity $\sigma$ associated to a local current-density $J_r$ (supported on a finite set of sites in the vicinity of $r$) reads: 
\begin{equation}\label{KuboDC}
\begin{split}
 &{\rm Re}[\sigma(\omega\to 0)]= \frac{ \pi \beta}{|\Lambda|}  \sum_{r' r} \sum_{m \neq m'}\frac{e^{-\beta E_{m'}}}{\mathcalnew{Z}} \times \\
 & \times  \langle E_{m'} |J_{r'+r}| E_{m} \rangle \langle E_{m} |J_{r'} |E_{m'} \rangle \, \delta_\eta\tonde{E_{m'}-E_{m}},
 \end{split}
\end{equation}
where $|\Lambda|$ is the system's volume, $\mathcalnew{Z}$ is the partition function at inverse temperature $\beta$, $|E_m\rangle$ the system's eigenstates and $\delta_\eta(x) = \pi^{-1}\eta/(x^2 +\eta^2)$ a regularized $\delta$-function. Since the set of LIOMs is complete, for any pair $m, m'$ there is a $\tilde I$ such that $\tilde I |E_m\rangle= \tilde I_m |E_m\rangle$ and $\tilde I |E_{m'}\rangle= \tilde I_{m'} |E_{m'}\rangle$ and $\tilde I_{m'}\neq \tilde I_{m}$. If the LIOMs are \emph{strictly local} (in the sense that their support is finite and compact, of size $\xi_{op}$), then for $r > \xi_{op}$, one of the two matrix elements
\begin{equation}\label{withcomm}
\begin{split}
 &\langle E_{m'} |{J_{r'}} |E_{m} \rangle = \frac{\langle E_{m'} |\quadre{J_{r'} ,\tilde I}|E_m \rangle}{ \tonde{\tilde I_m- \tilde I_{m'}} },\\
 &\langle E_{m'} |J_{r'+r} |E_m \rangle = \frac{\langle E_{m'} |\quadre{J_{r'+r} ,\tilde I}| E_m \rangle}{ \tonde{\tilde I_m- \tilde I_{m'}} }
 \end{split}
\end{equation}
 is exactly zero: in Eq.~(\ref{KuboDC}) the sum over $r$ is restricted to $r\lesssim \xi_{op}$. Furthermore, for any $m$ the sum over $m'$ is restricted to a finite set, since $J_{r'}|E_m \rangle$ can differ from $|E_m\rangle$ only in a finite number of LIOMs. As a consequence, in the thermodynamic limit, when $\eta\to 0$, the contribution to the $\delta$-function vanishes with probability one, and ${\rm Re}[\sigma(\omega=0)] =0$.
               For \emph{quasilocal} conserved quantities, the matrix elements $\langle m'| J_{r'}|m \rangle$ are not exactly zero also  for those eigenstates for which $\tilde{I}$ is supported at distance $x \xi_{op}$ from $r'$: they are instead exponentially small in $x$. Since there are also exponentially many states $m, m'$ which satisfy these criteria, some energy differences $E_m-E_m'$ in ($\ref{KuboDC}$) become exponentially small. The competition between the matrix elements and the energy denominators is however dominated by the exponential decay of the matrix elements with probability one: this is the key statement that guarantees the stability of the MBL phase as well as the existence of quasilocal $I_\alpha$. It follows that the conductivity remains zero also when quasilocality is properly taken into account.

     \item[(ii)] \emph{Anderson localization in configuration space: }      
     the exponential decay of the amplitudes \eqref{eq:QuasiLoc} can be interpreted as exponential localization in a space with sites labeled by $\mathcalnew{I}$, which  can be put in one to one correspondence with Fock states, see Sec. \ref{sec:PerturbationTheory}, or non-entangled product states in the spin case. Notice however that contrary to the Anderson model, in the many-body case an exponential decay of the amplitude in Fock space starting from a reference configuration does not imply $O(1)$ participation ratios for the wave function, due to the fast growth of the number of configurations with the distance from the reference configuration.
     
     %.  In the constructive procedures discussed in Sec. \ref{sec:analyticalConstruction}, the operators $I_\alpha$ are obtained as a weak deformation of the local, physical degrees of freedom diagonalizing the non-interacting part of the Hamiltonian (either the spin operators $\sigma^z_i$ or the occupation numbers of localized single particle states $n_\alpha$).  

 \item[(iii)] \emph{Area-law entanglement in highly excited states:} all the eigenstates of~\eqref{eq:EffMod} are product states of the $I_\alpha$; the bipartite entanglement entropy is thus area-law, as it is effectively contributed only by those $I_\alpha$ that are localized in the vicinity (within $\xi_{op}$) of the cut.

 \item[(iv)] \emph{Violation of the ETH:} as in conventional integrable systems, the locality of the conserved quantities implies that local memory of the initial condition is preserved at any time, thus preventing thermalization. Moreover, since the many-body eigenstates are simultaneous eigenstates of the $I_\alpha$ that are weak deformations of the non-interacting occupation numbers $n_\alpha$ (or of the spin operators $S^z_a$), it follows that the expectation value of the latter on exact eigenstates does not depart significantly from $\pm 1$. It is thus non-thermal and it fluctuates significantly between states that are close in energy, as such states differ by the eigenvalues of integrals of motion having a large overlap with the local operator under consideration. 
 
 \item[(v)] \emph{Absence of level repulsion:} 
 the absence of level repulsion arises because adjacent states in the spectrum typically differ by an extensive number of eigenvalues of the  $I_\alpha$, they are unable to hybridize and thus do not repel at the scale of the mean level spacing. 
 
 \item[(vi)] \emph{Slow growth of entanglement and slow relaxation:  }the interaction terms in~\eqref{eq:EffMod} induce dephasing between the conserved operators; given a pair of operators $I_\alpha, I_\beta$,
   their dephasing time $\tau$ scales with the inverse of the interaction between them, and thus it depends on the distance $l$ between the corresponding localization centers, $\tau \sim H_0^{-1} e^{l/ \xi_{int}}$ where $H_0$ is the typical interaction scale. It follows that at a given time $t$, only the degrees of freedom that are within the distance $l(t) \sim \xi_{int} \log (H_0 t)$ are dephased. The logarithmic growth of the bipartite entanglement follows from the fact that  the entanglement entropy produced at time $t$ (starting from a product state) is proportional to the volume of degrees of freedom that have dephased up to that time \cite{nanduri2014entanglement,Serbyn:2013he}. A similar argument implies that the expectation value of operators ${\rm O}$ with finite support decays as a power-law in time \cite{serbyn2014quenches}. Indeed, when expanded in the basis of the $I_\alpha, I^{\pm}_\alpha$, the operator reads
  \begin{equation}\label{eq:OpExpInv}
  {\rm O}=\overline{{\rm O}}+ {\rm O}_{osc}  ,
  \end{equation}
where ${\rm O}_{osc}$ contains $I^\pm_\alpha$ terms, while $\overline{{\rm O}}$ commutes with~\eqref{eq:EffMod}, and can be written as
 \begin{equation}\label{eq:TimeAv}
  \overline{{\rm O}}= \lim_{T \to \infty} \frac{1}{T} \int_0^T {\rm O}(t) \, dt = \sum_{\vec{\alpha}} \mathcalnew{C}_{\vec{\alpha}}\, {I_{\alpha_1 }I_{\alpha_2}\cdots I_{\alpha_n}} 
 \end{equation}
 where $\vec{\alpha}=(\alpha_1, \cdots, \alpha_n)$ (due to the quasilocality of the $I_\alpha$ and the locality of ${\rm O}$, the coefficients $\mathcalnew{C}_{\vec{\alpha}}$ have themselves an exponentially decaying structure). Let $ |\psi \rangle = \sum_{I} A_{I} \, | I \rangle $  be an initial state expanded in the basis of simultaneous eigenstates $|I \rangle$ of the $I_\alpha$. It is argued in \cite{serbyn2014quenches} that the second term in
\begin{equation}\label{eq:GenTimeEv}
\begin{split}
 \langle \psi(t) | {\rm O}| \psi(t) \rangle&= \sum_{I} |A_{I}|^2 \langle I | \overline{{\rm O}}| I \rangle\\
 &+ \sum_{I, J} A_{I} A^*_J e^{i(E_J -E_I)t} \langle J | {\rm O}_{osc}| I \rangle
 \end{split}
\end{equation}
 decays as a power law in time, due to the randomization of the relative phases in the eigenstates decomposition generated by the interactions of the $I_\alpha, I_\beta$. For example, let ${\rm O}_{osc}= 
 I^{+}_{\gamma}$, and consider for simplicity the case in which $|A_{I}|^2\sim 2^{-N}$ with $N$ the number of spins. For any state $| I \rangle$ with $i_\gamma=-1$, let $|J \rangle =|\tilde{I} \rangle$ be the state with $\tilde{i}_\gamma=+1$ and $\tilde{i}_\alpha= i_\alpha$ for $\alpha \neq \gamma$. The second term in \eqref{eq:GenTimeEv} becomes, using \eqref{eq:EffMod}, 
 \begin{equation}\label{eq:SecondTermAb}
  \frac{1}{2^N} \sum_{I: i_\gamma=-1} \text{exp} \quadre{2i \tonde{h_\gamma + \sum_{\beta} h_{\gamma \beta} \;i_\beta + \sum_{\beta, \delta} h_{\gamma \beta \delta} \; i_\beta i_\delta+\cdots}}.
 \end{equation}
  For any time $t$, \eqref{eq:SecondTermAb} can be splitted into the sum over quantum numbers $i_\alpha$ whose operators are localized within distance $l(t) \sim \xi_{int} \log (H_0 t)$ from $I_\gamma$, and the sum over quantum numbers of operators localized at larger distance. Any of the terms at the exponent involving quantum numbers in the first set is dephased at time $t$: \eqref{eq:SecondTermAb} is therefore proportional to a sum over $N(t)\sim 2^{2 l(t)}$ random phases with zero average, which scales as $\sim 1/\sqrt{N(t)}= (H_0 t)^{- \xi_{int} \log 2}$. 
  The expectation value~\eqref{eq:GenTimeEv} thus exhibits a power law relaxation to the constant (non-thermal) value
 $\sum_{I} |A_{I}|^2 \langle I | \overline{{\rm O}}| I \rangle$.
  
\end{itemize}
    
\section{Construction of conservation laws: analytic schemes}\label{sec:analyticalConstruction}
It was pointed out in~\cite{OganesyanHuseIntegrals} that, for an arbitrary many-body system with $N$ degrees of freedom, any bijection between the eigenstates and the $2^N$  binary strings $|i_{\alpha_1} i_{\alpha_2} \cdots  i_{\alpha_N} \rangle $ with $i_\alpha= \pm 1$ (or $i_\alpha \in \grafe{0,1}$) defines a complete set of conserved quantities $I_\alpha$ with binary spectrum $\grafe{i_\alpha}$. The latter are simply the operators whose quantum numbers $i_\alpha$ label the eigenstates. They can be obtained as $
 I_\alpha= \sum_{I'} | I' \rangle \langle I' |- |\overline{I'} \rangle \langle\overline{I'} |$,
assuming $i_\alpha= \pm 1$ and denoting with $| I' \rangle$ a many-body eigenstate with $i_\alpha= +1$, and with $|\overline{I'} \rangle$ the state with all quantum numbers equal except $i_\alpha=-1$ (see Eq. \eqref{tautau}). In \cite{OganesyanHuseIntegrals} it was conjectured that, for MBL systems, an optimal assignment exists that results in (quasi-)local operators $I_\alpha$. Although this has not led to a practical way to find LIOMs, that early work was instrumental for further development of the idea.

In this section, we analyze the constructive analytic recipes that have been proposed to construct the conserved operators starting from microscopic models. The discussion on numerical schemes is postponed to Sec. \ref{sec:numericalConstruction}.

\subsection{Infinite-time averages of local densities}\label{sec:TimeAv}
It was pointed out in~\cite{Chandran2014} that the infinite-time average of any local  operator ${\rm O}$ is a conserved quantity. If an underlying algebra of quasilocal operators $I_\alpha, I^\pm_{\alpha}$ exists, the average \eqref{eq:TimeAv} of operators with finite, compact support is itself quasilocal. For a spin Hamiltonian such as \eqref{eq:XXZ}, 
%\begin{equation}\label{eq:XXZ}
% H_{\text{XXZ}}= \sum_{a \in \Lambda} J \tonde{\sigma^x_a \sigma^x_{a+1}+\sigma^y_a \sigma^y_{a+1}} + J_z \sigma^z_a \sigma^z_{a+1}+ h_a \sigma^z_a=\sum_{a \in \Lambda}H_a,
%\end{equation}
an extensive set of LIOMs can be obtained as the infinite-time average of the local energy- or spin-densities operators, i.e. ${\rm O}_i \rightarrow S^z_i$ or ${\rm O}_i \rightarrow H_i \equiv J \tonde{S^x_i S^x_{i+1}+S^y_i S^y_{i+1}} + J_z S^z_i S^z_{i+1}+ h_i S^z_i$~\cite{Chandran2014}. The conserved operators which result from this procedure are not pseudospins, as the time-averaging does not preserve the spectrum. However, they have the advantage of being measurable in the following sense: for a chain with $N$ spins with ${\rm O}_i=S^z_i$, the coefficients in the expansion
\begin{equation}
 \overline{S}^z_i= \sum_{\vec{l}, \vec{\kappa}} \, M_{\vec{l}}^{\vec{\kappa}} \, S^{\kappa_1}_{l_1} \cdots S^{\kappa_n}_{l_n},
\end{equation}
with $\kappa_i \in \grafe{z, x, y}$, can be obtained measuring multi-spin correlations on a time-averaged density matrix $\overline{\rho}$,
\begin{equation}
 M_{\vec{l}}^{\vec{\kappa}}= 2^{-N} \text{Tr} \tonde{S^{\kappa_1}_{l_1} \cdots S^{\kappa_n}_{l_n} \, \overline{\rho}},
\end{equation}
where $\rho$ at $t=0$ describes the state with magnetization one at site $i$ and zero everywhere else, $\rho= 2^{-N} (1+ S^z_i)\otimes \prod_{k \neq i} {\bf 1}_k$.  The $\overline{S}^z_i$ thus provide information on the spreading of the spin through the infinite temperature ensemble. 

The alternative choice $\overline{{\rm O}}_i  \rightarrow \overline{H}_i$ is used in~\cite{Kim2014Lightcone} to derive a Lieb-Robinson bound for the information propagation in the MBL phase,
%\footnote{The Lieb-Robinson bounds limit the speed at which the information propagates under the dynamics given by quantum many-body systems with local interactions. They state that an effective speed of light exists, defining an effective lightcone, such that correlations outside the lightcone are exponentially suppressed in their distance ~\cite{LiebRobinson}.} 
with a logarithmic lightcone. The bound implies that the growth of the bipartite entanglement entropy is at most logarithmic. Its derivation relies on a stronger condition of quasilocality for the $\overline{H}_i$ (formulated in terms of averages rather than of typical values), requiring that there exists a constant $\xi$ such that
\begin{equation}\label{eq:NormDecay}
 \mathbb{E} \tonde{ \left\| \quadre{\overline{H}_i, {\rm O}} \right\|} \leq e^{-x/ \xi} ||{\rm O}||
\end{equation}
for any operator ${\rm O}$ whose support is at distance $x$ from the site $i$, where $\mathbb{E} \quadre{ \cdot}$ denotes the disorder average. %Numerical indications that the exponential decay~\eqref{eq:NormDecay} holds true for strong-enough disorder are reported in~\cite{Kim2014Lightcone}.  

\subsection{Perturbative dressing of the non-interacting occupation numbers}\label{sec:PerturbationTheory}
Consider the fermionic Hamiltonian \eqref{eq:FermHam}: in the absence of interactions $U=0$, the occupation numbers $n_\alpha$ associated to the single particle eigenstates $\phi_\alpha$ are mutually commuting, quasilocal conserved quantities. Their quasilocality follows directly from the spatial localization of the single particle wave-functions. Indeed, the operator $ c^{\dag}_i c_j$ contributes to the operator expansion:
\begin{equation}\label{eq:expsingle}
 n_\alpha= \sum_{i,j} \phi_{\alpha}^{*}(i) \phi_{\alpha}(j) c^{\dag}_i c_j
\end{equation}
with a weight $ \phi_{\alpha}^{*}(i) \phi_{\alpha}(j)$ which decays exponentially in the distance between its support (the sites $i,j$) and the localization center $ r_\alpha$ of $\phi_\alpha$. By truncating the sum \eqref{eq:expsingle} to terms with support only within a neighborhood of $m \xi$ of $ r_\alpha$ one obtains an operator whose commutator with the Hamiltonian vanishes up to exponentially small terms. As $m\to \infty$ the operator rapidly converges (in the operator norm) to the conserved density $n_\alpha$. It is natural to expect that for weak $U$, quasilocal operators that are conserved by the \emph{full} interacting Hamiltonian can be built as weakly dressed versions of the \eqref{eq:expsingle}. In order to confirm this expectation, it is necessary to show that the perturbation theory in operator space converges in the regime of parameters corresponding to weak scattering processes; this is argued in~\cite{rms-IOM} for a fermionic Hamiltonian on a $d$-dimensional lattice.

Before summarizing the main steps of the argument, we introduce the relevant energy scales in the problem. The convergence of the perturbative expansion is argued for a simplified version of the model~\eqref{eq:FermHam}, that is coarse-grained at the scale of the localization length $\xi$ of the single particle states $\phi_\alpha$. As in~\cite{Basko:2006hh}, the lattice $\Lambda$ is partitioned into volumina of size $\xi$ (henceforth called ``localization volumina''), and each single particle state is assigned to a volume according to the position of its localization center. An energy scale is associated to the localization volumina: it is the average energy-gap between the single-particle states belonging to it, $\delta_\xi= 1/ \nu \xi^d$ with $\nu$ the density of states. We assume $\delta_\xi \ll \mathcalnew{W}$ where $\mathcalnew{W}$ denotes the total width of the single particle spectrum. The interaction matrix elements $U_{\alpha \beta, \gamma \delta}$ are taken to be non-zero only if the corresponding single-particle states are in the same volume or in adjacent ones, and provided that 
\begin{equation}\label{eq:restriction}
 |E_\alpha- E_\delta|, |E_\beta- E_\gamma| \lesssim \delta_{\xi} \quad \text{or} \quad |E_\alpha- E_\gamma|, |E_\beta- E_\delta| \lesssim \delta_{\xi}.
\end{equation}
In this case, they are set equal to $U_{\alpha \beta, \gamma \delta}= \lambda \, \delta_\xi \, \eta_{\alpha \beta, \gamma \delta}$ where $\eta_{\alpha \beta, \gamma \delta}$ is a uniform variable in $\quadre{-1, 1}$ and $\lambda$ is a dimensionless constant measuring the interaction strength. Correlations between the single-particle energies $E_\alpha$ are neglected.

The setup for the construction is the following: an expansion is assumed for the conserved operators $I_\alpha$, of the form: 
\begin{equation}\label{eq:Ansatz}
 I_\alpha= n_\alpha + \sum_{N \geq 1} \sum_{\substack{\mathcalnew{I} \neq \mathcalnew{J}\\
 |\mathcalnew{I}|=N=|\mathcalnew{J}|}} \mathcalnew{A}^{(\alpha)}_{\mathcalnew{I,J}} \tonde{O_{\mathcalnew{I,J}}+ O^\dag_{\mathcalnew{I,J}}}
\end{equation}
where $\mathcalnew{I}=(\beta_1, \cdots, \beta_N)$ and $\mathcalnew{J}=(\gamma_1, \cdots, \gamma_N)$ are sets of indices labeling the single particle states, and 
\begin{equation}\label{eq:normOrd}
 O_{\mathcalnew{I,J}}= \prod_{\beta \in \mathcalnew{I}} c^\dag_\beta \prod_{\gamma \in \mathcalnew{J}} c_\gamma
\end{equation}
is a normal ordered operator (an ordering between the single particle indices is also assumed). The operator expansion~\eqref{eq:Ansatz} corresponds to a number operator dressed with strings of excitations. It is non-generic due to the constraint $\mathcalnew{I} \neq \mathcalnew{J}$, which is imposed to ensure that the coefficients $\mathcalnew{A}^{(\alpha)}_{\mathcalnew{I,J}}$ are \emph{uniquely} fixed by the conservation condition $\quadre{I_\alpha, H}=0$ (see the following discussion). The resulting coefficients depend on the interaction strength $\lambda$, and approach zero as $\lambda \to 0$. Whenever the operator expansion converges in norm, the Ansatz~\eqref{eq:Ansatz} defines a complete set of conserved quantities that are expected to be functionally independent at finite $\lambda$ (as they are for $\lambda=0$) and mutually commuting (the commutation relations $\quadre{I_\alpha, I_\beta}=0$ for $\alpha \neq \beta$ are not explicitly enforced, but assumed to hold due to the fact that the spectrum is almost-surely non degenerate.).

The goal is to argue that for sufficiently small $\lambda$ the expansion \eqref{eq:Ansatz} converges in probability, that is, that there is a $\lambda_c$ such that, for $\lambda < \lambda_c$ and for any $\epsilon>0$, the following condition is satisfied: 
\begin{equation}\label{eq:ConvCritPT}
 \lim_{R \to \infty}\mathbb{P} \tonde{\sum_{\substack{\mathcalnew{I} \neq \mathcalnew{J}\\
 %|\mathcalnew{I}|=|\mathcalnew{J}|\\
 r(\mathcalnew{I,J})>R}} |\mathcalnew{A}_{\mathcalnew{I,J}}^{(\alpha)}| <\epsilon}=1,
\end{equation}
where $r(\mathcalnew{I,J})$ is the maximal distance between the localization center of $\phi_\alpha$ and any of the states $\phi_\beta$ that are acted upon by $O_{\mathcalnew{I,J}}$. This ensures that the series defining the operator $I_\alpha$ converges almost surely, since $||O_{\mathcalnew{I},\mathcalnew{J}}||=1$ for all $\mathcalnew{I,J}$. The arguments for the convergence rely on a mapping to an equivalent single-particle problem, obtained imposing $\quadre{I_\alpha, H}=0$ and interpreting the resulting linear constraints for the amplitudes $\mathcalnew{A}^{(\alpha)}_{\mathcalnew{I,J}}$ as the equations for a single particle hopping in ``operator space''. The latter is a disordered lattice with sites labeled by the Fock indices $(\mathcalnew{I,J})$ and links determined by the conditions \eqref{eq:restriction} on the interactions. Within this formalism, the exponential decay of the coefficients $\mathcalnew{A}^{(\alpha)}_{\mathcalnew{I,J}}$ corresponds to the localization of the effective single particle in  operator space. At the operator level, it implies that the resulting operator $I_\alpha$ is quasilocal in the sense of \eqref{eq:QuasiLoc}. 

We now discuss in slightly more detail the main steps leading to \eqref{eq:ConvCritPT}.  We anticipate that the exponential convergence is proved within a ``forward approximation''~\cite{medina1992quantum, anderson1958absence, abou1973selfconsistent, pc2016forward} (henceforth FA), which in this setting boils down to replacing each amplitude in \eqref{eq:Ansatz} with its lowest order expansion in the strength of the interaction $\lambda$. This approach, although approximate, has the advantage of being explicit, and to allow one to obtain analytic expressions for relevant quantities characterizing the MBL phase \cite{rm2016remanent}. For an exact treatment that goes beyond this approximation, we refer the reader to Sec. \ref{sec:imbrie}.

\subsubsection{Formal perturbation theory and the problem of local resonances} 
When attempting to construct conserved quantities perturbatively, it is natural to consider the formal expansion 
\begin{equation}\label{eq:NaivePT}
 I_\alpha= n_\alpha + \sum_{n \geq 1} \lambda^n \Delta I_\alpha^{n},
\end{equation}
where $\Delta I_\alpha^{n}$ is determined recursively from $\Delta I_\alpha^{n-1}$, as the solution of the conservation equation 
\begin{equation}\label{eq:Cons}
 \quadre{U, \Delta I_\alpha^{n-1}}+ \quadre{H_0, \Delta I_\alpha^{n}}=0.
\end{equation} 
This equation can be solved for $ \Delta I_\alpha^{n}$ provided that the commutator with $H_0$ can be inverted; this requires that at any order $n$, the operator $\quadre{U, \Delta I_\alpha^{n-1}}$ does not belong to the kernel $K$ of the map $(\text{ad}\;{H_0})X= \quadre{H_0, X}$, which can be easily shown to be true for time-reversal symmetric Hamiltonians. The condition \eqref{eq:Cons} does not determine $\Delta I_\alpha^{n}$ uniquely: indeed, arbitrary terms $\Delta K_\alpha^{n}$ belonging to $K$ can be added to $\Delta I_\alpha^{n}$, and the resulting operator $\Delta I_\alpha^{n}+ \Delta K_\alpha^{n}$ would still satisfy \eqref{eq:Cons}. We may thus write
\begin{equation}\label{eq:SolPT}
\begin{split}
 \Delta I^{n}_\alpha &= i \lim_{\eta \to 0} \int_0^\infty d\tau e^{-\eta \tau} e^{i H_0 \tau} \quadre{U, \Delta I_\alpha^{n-1}} e^{-i H_0 \tau}+ \Delta K_\alpha^{n}\\
 &= \Delta J^{n}_\alpha + \Delta K^n_\alpha,
 \end{split}
\end{equation}
where the first term in~\eqref{eq:SolPT} belongs to the space spanned by the normal ordered operators~\eqref{eq:normOrd} with $\mathcalnew{I} \neq \mathcalnew{J}$ (which is the orthogonal complement of $K$), while $\Delta K_\alpha^{n} \in K$. It is shown in \cite{rms-IOM} that there exists a unique choice of $\Delta K_\alpha^{n} \in K$ that ensures that the spectrum of \eqref{eq:NaivePT} is binary, i.e.  $I^2_\alpha=I_\alpha$, at the given perturbative order. This term can be written as
\begin{equation}\label{ExpDiag}
  \Delta K^{(n)}_\alpha=\tonde{1-2 n_\alpha} \quadre{\sum_{m=1}^{n-1} \Delta I_\alpha^{m} \Delta I_\alpha^{n-m} + \grafe{n_\alpha-\frac{1}{2}, \Delta J_\alpha^{n} }}.
 \end{equation}
Thus, the formal perturbative recipe uniquely defines a set of conserved number operators obtained summing \eqref{eq:SolPT}. Let us now address the problem of convergence and quasilocality.

Despite the perturbative equations are solvable at any finite order, the resulting series~\eqref{eq:NaivePT} diverges almost surely, due to occurrence of terms with large norm that, even if rare, appear \emph{repeatedly} in~\eqref{eq:NaivePT}, at any order in $\lambda$, leading to a divergence. These are the same ``trivial'' divergences affecting the perturbation theory for the self energies discussed in P.~W.~Anderson's work~\cite{anderson1958absence}; they signal the presence of \emph{rare resonances} between almost degenerate states of the unperturbed Hamiltonian, that are strongly hybridized by the hopping/interactions. The presence of isolated resonances does not necessarily entail delocalization, as long as they do not proliferate in space; in \cite{anderson1958absence}, the corresponding divergences are addressed by means of the ``multiple scattering technique'', which consists in a re-summing the subsequences containing repeating terms. As a result, a modified perturbative expansion is obtained as a sum over non-repeating terms, with self energy corrections arising from the re-summed subsequences. The latter are subsequently neglected in the so called forward or ``upper limit'' approximation, which leads to an overestimate of the critical value of the disorder for the onset of delocalization. The arguments given in~\cite{rms-IOM} for the convergence of \eqref{eq:Ansatz} are exactly at this level of approximation; however, to recover an analogous expansion in terms of non-repeating terms, it turns out to be technically convenient to drop the terms $\Delta K_\alpha^{n}$ in~\eqref{eq:SolPT}. The resulting operator $I_\alpha- n_\alpha$ belongs then to the orthogonal complement of $K$, and so it admits the expansion~\eqref{eq:Ansatz}. This choice guarantees that the coefficients are uniquely fixed \emph{just} by imposing $\quadre{H, I_\alpha}=0$, without any additional requirement. The drawback is that the ansatz~\eqref{eq:Ansatz} does not define an operator with binary spectrum; however, it is expected that the statement on the finiteness of the radius of convergence is not spoiled once the neglected terms are reinserted and the ``normalization'' of \eqref{eq:Ansatz} is imposed.

    \begin{figure*}[!htb]
   %\captionsetup[subfigure]{labelformat=empty}
  \centering
        \subfigure[]{\includegraphics[scale=0.35]{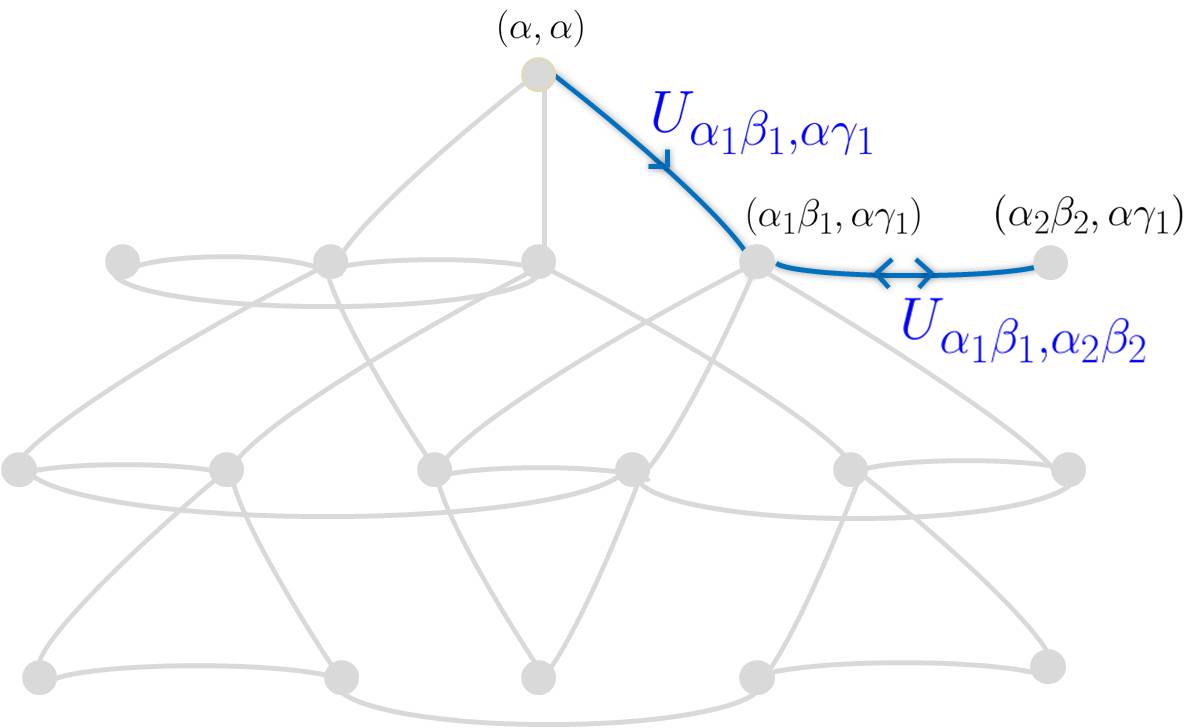}}\quad
        \subfigure{\includegraphics[scale=0.35]{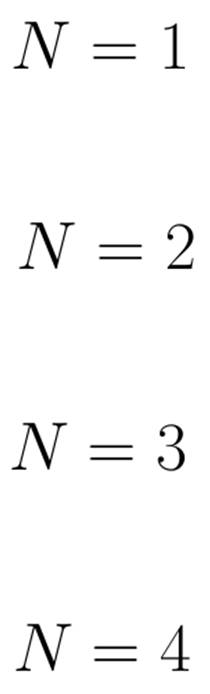}}\quad
        \subfigure[]{\includegraphics[scale=0.35]{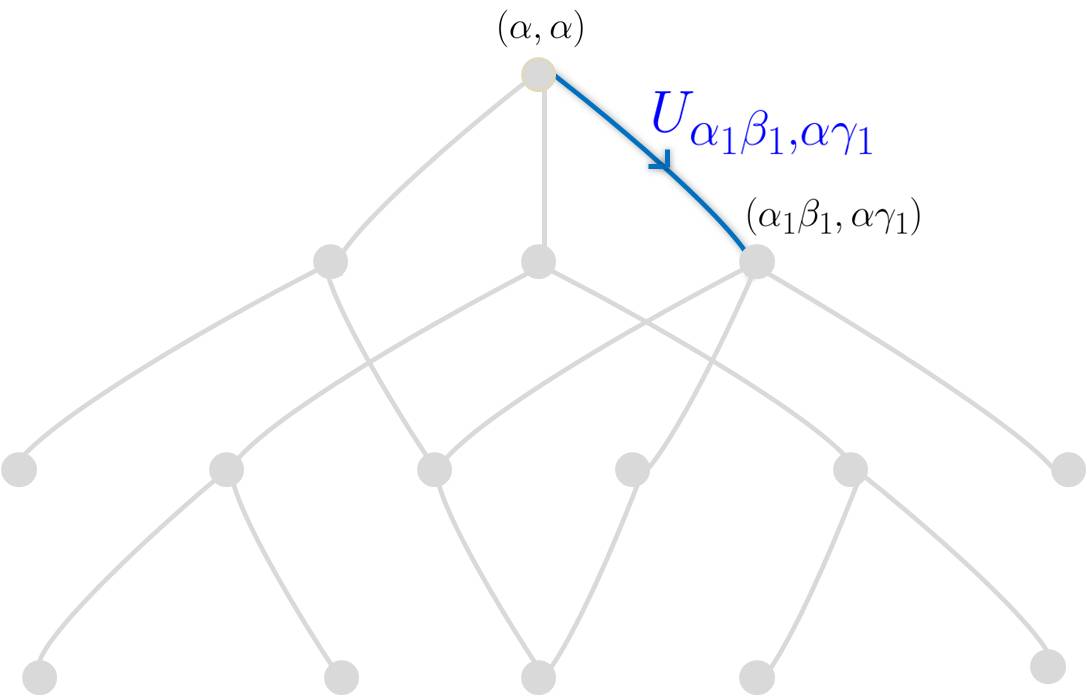}}
        \caption{Structure of the operator lattice before (a) and after (c) making the forward approximation. Vertices correspond to Fock indices ($\mathcalnew{I,J}$); links are drawn between index pairs, which are connected by the interaction $U$. }
  \label{fig:lattice}
    \end{figure*}

\subsubsection{Mapping to a single-particle problem and forward approximation} 
The condition  $\quadre{I_\alpha, H}=0$ translates into a set of linear constraints for the amplitudes $\mathcalnew{A}^{(\alpha)}_{\mathcalnew{I,J}}$, cfr. Eq.(40) in \cite{rms-IOM}. They define an effective non-hermitian hopping problem on a lattice with sites labeled by $(\mathcalnew{I,J})$, on-site disorder $\mathcalnew{E}_\mathcalnew{I,J}= \sum_{n=1}^N (E_{\beta_n}- E_{\gamma_n})$ and hopping given by the interactions: precisely, two sites ($\mathcalnew{I,J}$) and $(\mathcalnew{I',J'})$ are connected if their index sets differ by indices belonging to a non-zero matrix element $U_{\alpha \beta, \gamma \delta}$. A pictorial representation of the resulting lattice is given in Fig. \ref{fig:lattice}. The main feature is its hierarchical structure: sites are organized into generations $N$ according to the length of the index sets, and the hopping is only within the same generation or from one generation to the next; in the latter case, it is unidirectional (hence the non-hermiticity of the hopping-problem). The typical connectivity is estimated accounting for the space (particles need to be in the same or in an adjacent localization volume) and phase-space restrictions \eqref{eq:restriction} imposed to the interaction matrix elements. For hoppings $U_{\alpha \beta, \gamma \delta}$ between consecutive generations, it scales as
\begin{equation}\label{eq:connectivity}
\mathcalnew{K} =4 \frac{\mathcalnew{W}}{\delta_\xi}.
\end{equation}
Such hoppings indeed require a particle (or hole) $\alpha$ in $(\mathcalnew{I,J})$ to scatter to the state $\gamma$ within the localization volume that is closer in energy (above or below $\alpha$), while another particle-hole pair of levels $(\beta,\delta)$ with adjacent energies is created. The particle $\beta$ can be chosen in $\mathcalnew{W}/{\delta_\xi}$ ways, and there are two choices for $\gamma$ and $\delta$, respectively, hence \eqref{eq:connectivity}. 
In contrast, hoppings to sites of the same generation correspond to processes where each member of a pair of particles (or holes) scatter to one of the two closest energy levels: there are order $O(1)$  possible final states to which a fixed given pair can decay. 

\begin{figure*}[ht!]
     \centering
%\captionsetup[subfigure]{labelformat=empty}
  \subfigure[]{%
          \includegraphics[width=0.5\textwidth]{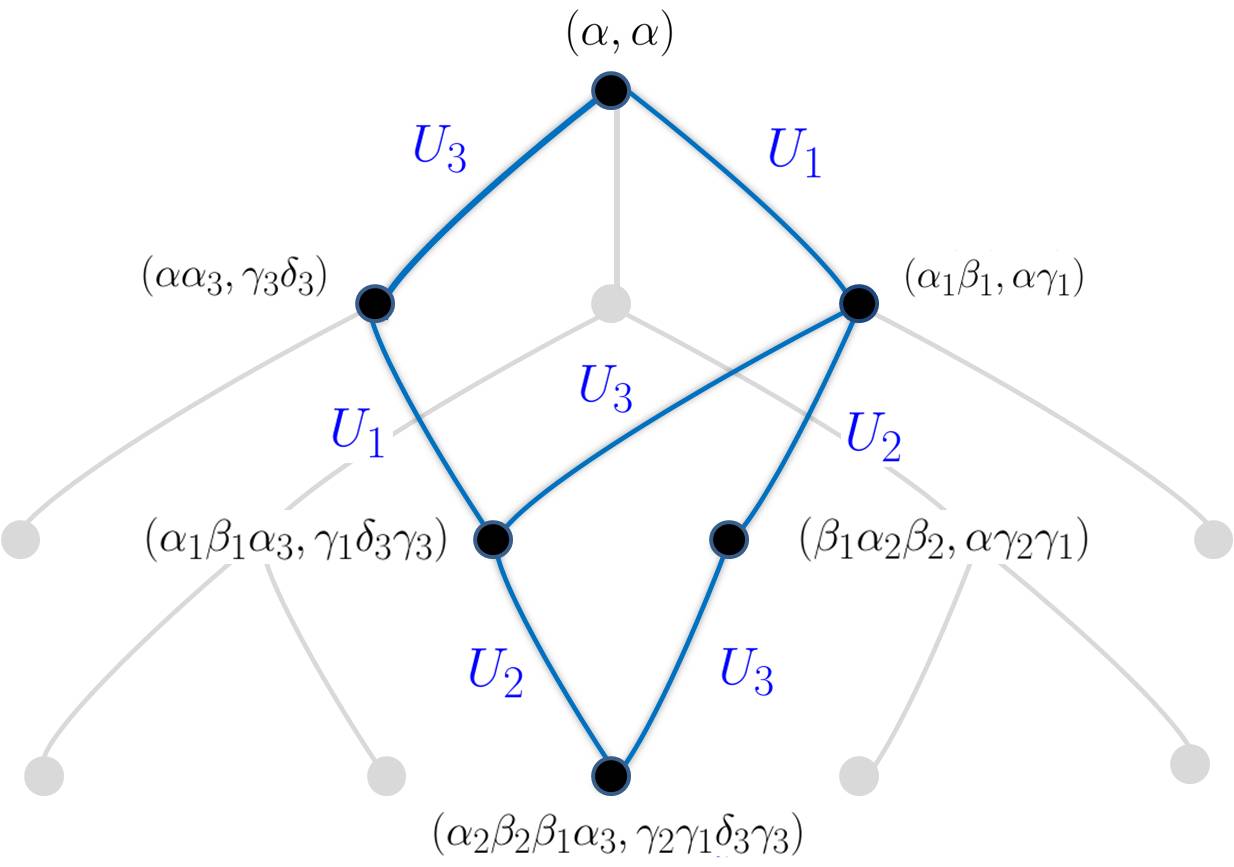}
        } \\%  ------- End of the first row ----------------------%
            \subfigure[]{%
           %\label{fig:path1}
           \includegraphics[width=0.3\textwidth]{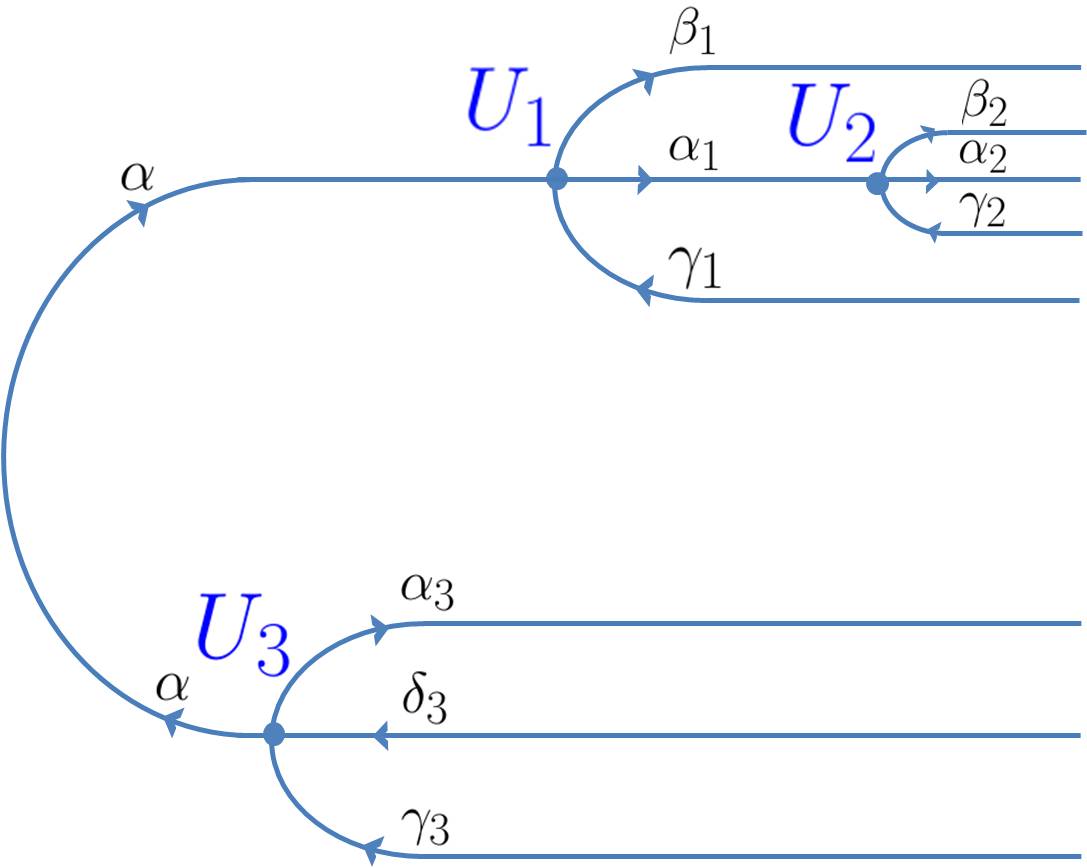}
        } %  ------- End of the first row ----------------------%
        \subfigure[]{%
            %\label{fig:path2}
            \includegraphics[width=0.3\textwidth]{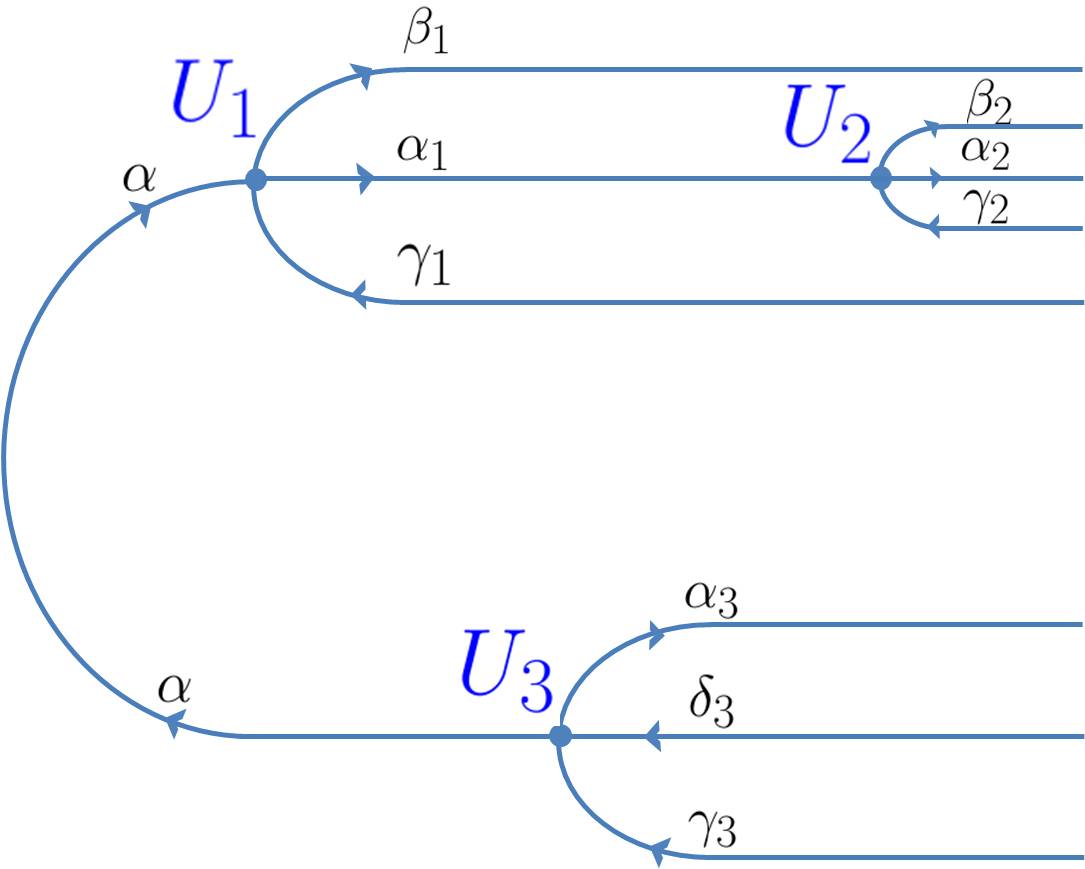}}%
        \subfigure[]{%
          % \label{fig:Paths2}
            \includegraphics[width=0.3\textwidth]{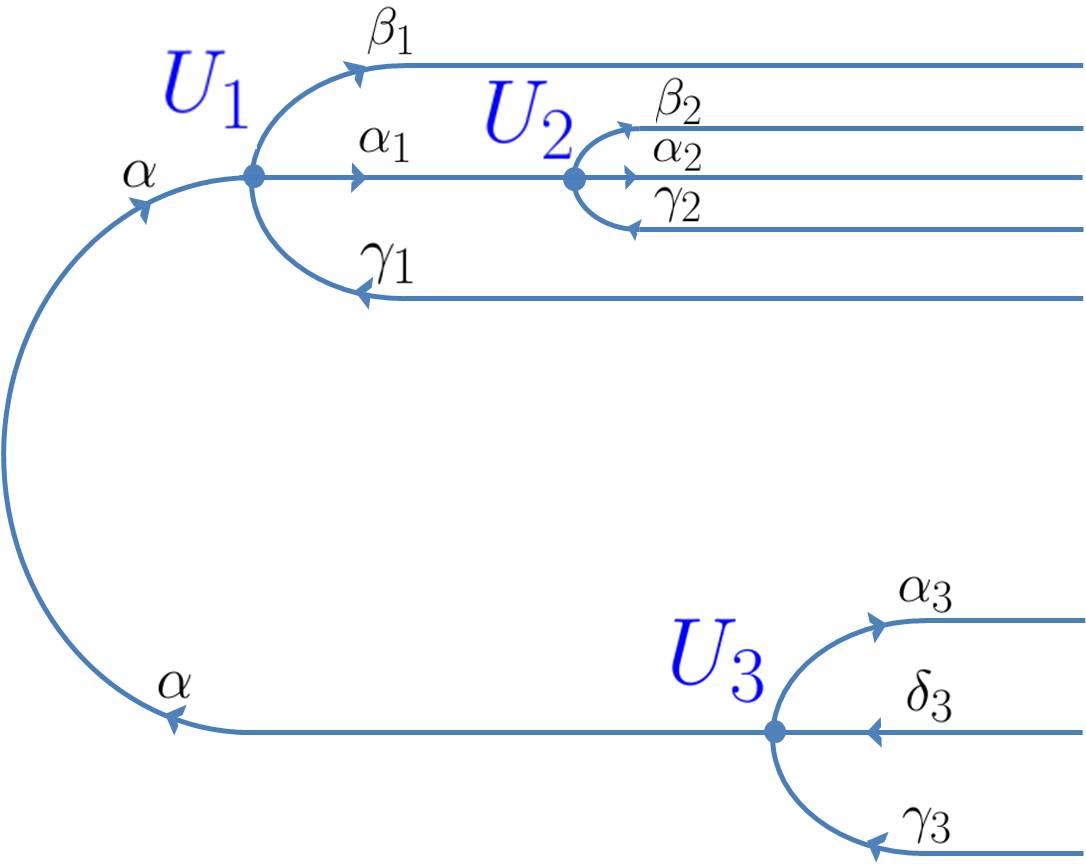}}%
\caption{Directed paths in the operator lattice and associated scattering graphs: the sites along the path are intermediate states of the graph, the hopping terms correspond to vertices $U_{\alpha_1 \alpha_2, \beta_1 \beta_2}$ in the graph (denoted with the simplified notation $U_1$, $U_2$, $U_3$), and the energy $\mathcalnew{E}_{\mathcalnew{I,J}}$ is the sum of the energy differences $\mathcalnew{E}_{\alpha_1 \alpha_2, \beta_1 \beta_2}= E_{\alpha_1}+E_{\alpha_2}-E_{\beta_1} -E_{\beta_2}$ associated with all preceding scatterings. The three highlighted paths form loops in the lattice, associated to graphs  differing only in the order in which the interactions $U_1$, $U_2$, $U_3$ act.}\label{fig:Paths2}
  \end{figure*}

 The forward approximation consists in neglecting the links connecting sites in the same generation of the lattice, which is justified provided~\eqref{eq:connectivity} is large. This significantly simplifies the lattice topology, as some sites are eliminated (the corresponding amplitudes in \eqref{eq:Ansatz} approximated to zero), see Fig.~\ref{fig:lattice}. The equations for the amplitudes on the retained sites become recursive equations for increasing generations, with initial condition $\mathcalnew{A}^{(\alpha)}_{\alpha,\alpha}=1$; this allows to derive a closed expression for the amplitude at the sites $(\mathcalnew{I,J})$ as a sum over all directed, non-repeating  paths~\cite{pc2016forward} from the root ($\alpha,\alpha$) to the given site:
\begin{equation}\label{eq:PathAmplitude}
\begin{split}
 \mathcalnew{A}^{(\alpha)}_{\mathcalnew{I,J}} &=  \sum_{\substack{\text{directed paths}\\
 (\alpha, \alpha) \to (\mathcalnew{I, J)}}} (-1)^{\sigma_{\text{path}}} \prod_{i=1}^{N-1} \frac{\lambda \, \eta_{\alpha_i \beta_i, \gamma_i \delta_i} \, \delta_\xi}{\sum_{k=1}^i \mathcalnew{E}_{\alpha_k \beta_k, \gamma_k \delta_k}}\\
 &\equiv \sum_{\substack{\text{directed paths}\\
 (\alpha, \alpha) \to (\mathcalnew{I, J)}}} \omega^{(N)}_{\text{path}},
 \end{split}
\end{equation}
where the factor $(-1)^{\sigma_{\text{path}}}$ is a global sign. The expression \eqref{eq:PathAmplitude} is of order $\lambda^{N-1}$, which is the lowest possible order for amplitudes of operators involving $2N$ particle-hole indices. Thus, for any $N$, any term in \eqref{eq:Ansatz} that is of order $O(\lambda^{N})$ or smaller is set to zero. The remaining terms satisfy $r(\mathcalnew{I,J}) \leq N \xi$ due to the locality of the interaction; it follows that within this approximation $r(\mathcalnew{I,J})$ in \eqref{eq:ConvCritPT} can be replaced by $N$, so that the convergence is controlled by the generation $N$. By eliminating intra-generation hoppings, the FA addresses the problem of arbitrary repetitions of small denominators; further insights on its meaning is derived from the diagrammatic representation of paths illustrated in  Fig. \ref{fig:Paths2}. The expansion in directed paths corresponds in fact to selecting only the processes where at each vertex an additional particle-hole pair is created; this is similar to the imaginary self consistent Born approximation exploited in \cite{Basko:2006hh}.

\subsubsection{Issue of factorials and re-summation of correlated graphs} 
Within the FA, the number of sites $(\mathcalnew{I,J})$ at distance $N$ from the root scales exponentially, as $\sim \mathcalnew{K}^N$. Moreover, the \emph{typical} path weights in \eqref{eq:PathAmplitude} decay exponentially with $N$ for small enough $\lambda$; thus, one might expect that a transition occurs, arising from the competition between these two exponentials. Although this will turn out to be precisely the case, an intermediate step has to be performed. The sum \eqref{eq:PathAmplitude} is indeed over a number of terms which grows faster-than-exponential with $N$, since to any given path of length $N$, one can associate a class of $O(\sim N!)$ other paths corresponding to graphs sharing the same interaction vertexes, but occurring in different order, see Fig. \ref{fig:Paths2}. Due to the structure of the energy denominators, each such graph has a different weight contributing to \eqref{eq:PathAmplitude}. A crucial step in arguing the exponential decay of \eqref{eq:PathAmplitude} consists in showing that this factorial is compensated, as cancellations occur between the different terms that are due to their statistical correlations. For instance, in the case of Fig. \ref{fig:Paths2} it holds
\begin{equation}
\begin{split}
 &\quadre{\frac{1}{\mathcalnew{E}_3(\mathcalnew{E}_3+\mathcalnew{E}_1)}+\frac{1}{\mathcalnew{E}_1(\mathcalnew{E}_1+\mathcalnew{E}_3)}+\frac{1}{\mathcalnew{E}_1(\mathcalnew{E}_1+\mathcalnew{E}_2)}}\frac{1}{\mathcalnew{E}_1+\mathcalnew{E}_2+\mathcalnew{E}_3}=\\
 &\frac{1}{\mathcalnew{E}_3 \, \mathcalnew{E}_1\,(\mathcalnew{E}_1+\mathcalnew{E}_2) },
 \end{split}
\end{equation}
where $\mathcalnew{E}_i$ is the linear combination of single particle energies associated to the interaction vertex $U_i$. 
This problem of factorials can be dealt systematically, giving an integral representation of the sum over path-weights differing by permutations of the interactions, and optimizing on the sequence of pole integrations that can be performed on the representation. As a result, the sum over the factorially-many paths is explicitly rewritten as a sum over only exponentially-many terms $\tilde{\omega}^{(N)}_{\text{path}}=O(\lambda^N)$ having a similar structure as \eqref{eq:PathAmplitude}, termed ``effective paths'': \begin{equation}\label{eq:PathAmplitudeR}
 \mathcalnew{A}^{(\alpha)}_{\mathcalnew{I,J}} = \sum_{\substack{\text{effective paths}\\
 (\alpha, \alpha) \to (\mathcalnew{I, J)}}} \tilde{\omega}^{(N)}_{\text{path}}.
\end{equation}

\subsubsection{The estimate of the convergence radius} For large $N$, the generating function of the random variables $\tilde{\omega}^{(N)}_{\text{path}}$ (analogous to the quantity in \eqref{(11)}) is computed by means of a transfer matrix technique, accounting for the correlations of energies in the denominators. The 
tail of the distribution of the (effective)-path weights is obtained by inverse Fourier transform, within a saddle point approximation. It is shown that the (effective)-path weights are heavy-tailed distributed, so that the  sum \eqref{eq:PathAmplitudeR} is dominated by the maximal summand. The typical value of \eqref{eq:PathAmplitudeR} is then estimated with a calculation that is close to the one performed in \cite{altshuler1997quasiparticle} for the Hamiltonian \eqref{eq:Hamiltonian} on a Bethe lattice: neglecting the residual correlations between the effective paths, one finds that for $z<1$
\begin{equation}\label{eq:FullProb}
\begin{split}
 &\log \mathbb{P} \tonde{\sum_{\substack{\mathcalnew{I} \neq \mathcalnew{J}\\
 %|\mathcalnew{I}|=|\mathcalnew{J}|\\
 |\mathcalnew{I}|=N =|\mathcalnew{J}|}} |\mathcalnew{A}_{\mathcalnew{I,J}}^{(\alpha)}| <z^N} \approx - \mathcalnew{N}_N \, \mathbb{P} \tonde{|\tilde{\omega}^{(N)}_{\text{path}}| >z^N}\\
 &\approx -\text{exp}\tonde{N \log \mathcalnew{G}(\lambda, z, \mathcalnew{K}) + o(N) },
 \end{split}
\end{equation}
where the function $\mathcalnew{G}(\lambda, z, \mathcalnew{K})$ combines the exponential growth (with $N$) of the number  $\mathcalnew{N}_N$ of effective paths of length $N$ with the exponential decay of the large-deviation probability $\mathbb{P} \tonde{|\tilde{\omega}^{(N)}_{\text{path}}| >z^N}$.  
The number $\mathcalnew{N}_N$ is determined by means of a combinatorial estimate of the number of diagrams representing the scattering processes with fixed final state, which accounts for the locality of the interactions by enforcing that only particle-hole pairs in nearby localization volumina can be involved in the same interaction vertex. Imposing $\mathcalnew{G}(\lambda, z, \mathcalnew{K})<1$ for $z \to 1$, the following condition for the convergence radius is obtained:
\begin{equation}\label{eq:LambdaCrit}
 \lambda_c= \frac{C}{\nu_{F} (1-\nu_{F})} \frac{1}{\mathcalnew{K} \log \mathcalnew{K}},
\end{equation}
  with $C$ a numerical constant estimated to be $18<C<37$, $\nu_F$ the filling fraction and $\mathcalnew{K}$ given in \eqref{eq:connectivity}. The many-body nature of the processes involved is evident in the fact that $\lambda_c\to\infty$ when $\nu_F\to 0$, while the divergence at $\nu_F\to 1 $ is a Fermi blockade effect. Thus, we conclude that for $\lambda< \lambda_c$ and within the set of approximations made, quasilocal conserved quantities for the interacting Hamiltonian \eqref{eq:FermHam} can be constructed as weak perturbations of the non-interacting occupation numbers $n_\alpha$. 

The estimate \eqref{eq:LambdaCrit} is in agreement with the  perturbative results in \cite{Basko:2006hh, gornyi2005interacting}, where a finite temperature transition is predicted. Indeed, so far the convergence of the construction was analyzed at the operator level, with no assumption made on the occupation of the single-particle energy levels. When acting on some typical many-body states at temperature $T$, some terms in \eqref{eq:Ansatz} get annihilated when attempting to create particles on occupied states or holes on already empty states. This translates into a reduction of the phase space associated to the decay processes, which in turns implies that $\mathcalnew{K}  \to \mathcalnew{K}_{\rm eff}\sim T/\delta_\xi$: substitution into \eqref{eq:LambdaCrit} gives the condition for the critical temperature in \cite{Basko:2006hh, gornyi2005interacting}. The agreement is not surprising, as the class of diagrams that are statistically analyzed is the same as within the forward approximation. It is however argued in \cite{de2016absence} (see also the discussion in \cite{rms-IOM}) that this scenario of a finite-temperature transition becomes unstable beyond this set of approximations, due to rare fluctuations within typical, putatively localized states, that are argued to restore ergodicity in the long time limit. We comment further on this point in Sec.\ref{sec:Discussion}.

\subsection{Nonperturbative construction of Local Integrals of Motion}\label{sec:imbrie}
%The arguments in the previous section entail that for $\lambda$ small enough the probability of collective, high-order resonances between unperturbed degrees of freedom decay exponentially fast in their distance, cfr. Eq. \eqref{eq:FullProb}. 
We now come to the discussion of the nonperturbative construction of local integrals of motion given in  \cite{imbrie2014many}. The work contains a rigorously proof of many-body localization for the spin chain \eqref{eq:HamImbrie} (which we denote simply with $H$ in the following) in its strongly disordered limit, following from a physically reasonable assumption on level statistics. The proof proceeds by constructing a sequence of quasilocal unitary rotations that diagonalize the Hamiltonian; the conserved quantities are obtained as a by-product of the diagonalization procedure. We begin by reviewing the construction in 
\cite{imbrie2014many,Imbrie2016c}, and subsequently discuss its implications for local integrals of motion.

As the spin flip interactions in \eqref{eq:HamImbrie} are of order $\gamma \ll 1$, the Hamiltonian is effectively in the strong-disorder regime. In the present discussion the focus is on the disorder in the random magnetic field $h_i$: the randomness is added to the other two terms in the Hamiltonian as it helps  with level separation arguments, but should not be essential to the physics of this model. For small $\gamma$, the Hamiltonian is close to one that is diagonal in the basis given by the tensor products of $S_i^z$ eigenstates. The MBL transition in this model can be thought of as a many-body version of what happens for a single spin: when $\Lambda$ is a single site, the Hamiltonian reduces to
\begin{equation}\label{(1.1)}
 \left( \begin{array}{cc}
h &\gamma \\\gamma & -h  \end{array} \right),
\end{equation}
having eigenvectors close to $(1,0)$ and $(0,1)$ for small $\gamma$, whereas for large $\gamma$, they are close to the fully hybridized vectors $(1,1)$ and $(1,-1)$. The result in \cite{imbrie2014many} extends this picture to the many-body Hamiltonian, demonstrating (under the assumption on level statistics) the existence of an MBL phase where the eigenstates of $H$ resemble the basis vectors, \textit{i.e.} they are largely unhybridized. 
Precisely, the following statement is proven:
\begin{theorem} \label{thm:1}
Assume that for some $\nu, C$, the eigenvalues of $H$ in boxes of size $n$ satisfy
\begin{equation}\label{(2)}
P\left(\min_{\alpha \ne \beta} |E_{\alpha} - E_{\beta}| < \delta\right) \leq \delta^{\nu} C^n, 
\end{equation}
for all $\delta>0$ and all n.
Then there exists a $\kappa > 0$ such that for $\gamma$ sufficiently small,
 \begin{equation}\label{(1.4)}
\mathbb{E}\, \normalfont{\text{Av}}_\alpha \left| \langle S^z_0 \rangle_\alpha\right| = 1 - O(\gamma^\kappa), 
\end{equation}
for all $\Lambda$. Here $\mathbb{E}$ denotes the disorder average,  $\langle \cdot \rangle_\alpha$ denotes the expectation in an eigenstate $\alpha$ of $H$, and $\mathrm{Av}_{\alpha}$ denotes an average over $\alpha$. 
\end{theorem}
The normalized average $\text{Av}_\alpha$ over the eigenstates of $H$ can be taken with thermal weights $\text{(const)}\exp(-\beta E_\alpha)$, which at infinite temperature becomes a uniform weighting.
Thus with high probability, most states have the property that the expectation of $S_0^z$ is close to $+1$ or $-1$, just like the basis vectors. 
This would contrast with a thermalized phase, with states approximating thermal ensembles, as expected from the Eigenstate Thermalization Hypothesis recalled in Sec. \ref{sec:phenomenology}. Indeed, at infinite temperature thermalization would imply that eigenstate expectations of $S_0^z$ would go to zero in the infinite volume limit; the bound (\ref{(1.4)}) implies a failure of thermalization, a key feature of the MBL phase.

The level-statistics assumption (\ref{(2)}) specifies that with high probability the minimum level spacing should be no smaller than an exponential in the volume. Physically, one expects to see (\ref{(2)})
satisfied with $\nu = 1$ in a localized phase (Poisson statistics) or with $\nu > 1$ in a thermalized phase (repulsive statistics). Although these bounds are expected to hold, the tools for proving them are not yet available in the many-body context (see  \cite{Imbrie2016a} for a potentially useful approach in the one-body context). We discuss in the following the main steps in the diagonalization procedure.

\subsubsection{A multiscale procedure to diagonalize an MBL Hamiltonian}
The proof proceeds by diagonalizing $H$ through successive elimination of low-order off-diagonal terms as in Newton's method.
Elements of the tensor product basis can be labeled by classical spin configurations $\sigma = \{\sigma_i\}_{i\in\Lambda}$, with $\sigma_i = \pm 1$ indicating the eigenvalue of $S_i^z$.
Initially, the only off-diagonal term is $\gamma_i S^x_i$, which is local. This operator flips $\sigma_i \rightarrow -\sigma_i$. 
Let  $\sigma^{(i)}$ denote the result of flipping the spin at $i$ in the spin configuration $\sigma$. Then the spin flip produces a change in energy
\begin{equation}\label{(5)}
\Delta E_i \equiv E(\sigma) - E (\sigma^{(i)}) = 2 \sigma_i (h_i + J_i \sigma_{i + 1} + J_{i - 1} \sigma_{i - 1}). 
\end{equation}
The site $i$ is said to be resonant if $|\Delta E_i|<\varepsilon \equiv \gamma^{1/20}$ for at least one choice of $\sigma_{i-1}, \sigma_{i+1}$. 
Due to the small energy denominator, resonant sites may require a rotation that is far from the identity, and perturbation theory is not useful. A site is resonant with probability $\sim 4\varepsilon$, so resonant sites form a dilute set.
 
Ignore the resonant sites, for the moment. A rotation can be designed that eliminates the off-diagonal terms $J(i) \equiv \gamma_i S_i^{x}$ in the Hamiltonian.
Let $H=H_0 + \mathcalnew{J}$, where $H_0$ contains the diagonal part of $H$ (first and third terms of (\ref{eq:HamImbrie})) and $\mathcalnew{J}$ contains the off-diagonal part (second term of (\ref{eq:HamImbrie})). 
Then define an antisymmetric matrix
\begin{equation}\label{(6)}
A \equiv \sum_{i} A(i), \text{ where }A(i)_{\sigma \sigma^{(i)}} = \frac{J(i)_{\sigma \sigma^{(i)}}}{E_\sigma-E_{\sigma^{(i)}}}.
\end{equation}
This is a local operator, which can be used to generate a basis change $ e^{-A}$.
The result is a rotated (or
renormalized) Hamiltonian:
\begin{equation}\label{(7)}
H^{(1)}=e^{A} H e^{-A}.
\end{equation}
This rotation is correct to first order in perturbation theory, because 
\begin{equation}\label{(8)}
[A,H_0] = -\mathcalnew{J};
\end{equation}
this enables the cancellation of the off-diagonal terms to leading order:
\begin{align}\label{(9)}
H^{(1)}
 &=  e^AHe^{-A} = H+[A,H]+ \frac{[A,[A,H]]}{2!} + \ldots 
  \nonumber
\\& =  H_0+\sum_{n=1}^\infty\left(\frac{1}{n!} - \frac{1}{(n+1)!}\right)(\text{ad}\,A)^n\mathcalnew{J}
\nonumber
\\& \equiv  H_0+J^{(1)}.
\end{align}
Here $(\text{ad}\,A)B\equiv [A,B]$. See \cite{Datta1996} for a similar construction. See also \cite{Imbrie2016b}, where this method is used to diagonalize the Anderson model Hamiltonian with weak hopping.
The new interaction
$J^{(1)}$ contains terms that are at least second order in $\gamma$.
Note that
$A(i)$ commutes with $A(j)$ or $J(j)$ if $|i-j|>1$.
One can write $J^{(1)}=\sum_g J^{(1)}(g)$, where $g$ is
a connected graph specifying a nonvanishing term of  $(\text{ad}\,A)^n J$. A graph involving $\ell$ spin flips has $\ell-1$ energy denominators and is bounded by $\gamma(\gamma/\varepsilon)^{\ell-1}$. 
Thus the size of terms in $J^{(1)}$ decay exponentially with their range: the interaction is quasilocal in the sense of \eqref{eq:QuasiLoc}.
    
If there are resonant sites present, then one may restrict the sum over $i$ in (\ref{(6)}) to nonresonant sites, thereby avoiding any small denominator $E_\sigma-E_{\sigma^{(i)}}$. 
Then the rotation (\ref{(7)}) eliminates terms of order $\gamma$ in the nonresonant region. To handle the resonant region,
define resonant blocks by taking connected components of the set of resonant sites. 
As in quasidegenerate perturbation theory, one performs
exact rotations in resonant blocks to diagonalize the Hamiltonian there. 
These rotations can be far from the identity, as resonances can lead to significant hybridization of spins. 
Thus, the notion of quasilocality has to be broadened to include the possibility of a dilute set of sites where a nontrivial basis change is required. 

Introducing a sequence of length scales $L_k =(15/8)^k$, the procedure
continues by defining the $k^{\text{th}}$ rotated Hamiltonian $H^{(k)}$, which has off-diagonal terms eliminated up through order $\gamma^{L_k}$. 
(As in Newton's method, the iteration converges at a rate that is close to quadratic.) 
In each step, the diagonal elements of $H^{(k)}$ are renormalized by interactions up to the $k^{\rm th}$ scale; they are denoted $E^{(k)}_\sigma$. In step $k \ge 2$, one needs to work with a graph-based notion of resonance. By definition, a graph
 $g$ is resonant if 
 \begin{equation}\label{(10a)}
 A^{(k+1)}_{\sigma \tilde{\sigma}}(g) \equiv \frac{J^{(k)}_{\sigma\tilde{\sigma}}(g)}{E^{(k)}_\sigma-E^{(k)}_{\tilde{\sigma}}}  
 \end{equation}
is greater than $(\gamma/\varepsilon)^{|g|}$ in magnitude. Here
$|g|$ denotes the number of spin flips in $g$.
For nonresonant graphs, the generator (\ref{(10a)}) is used to generate the next rotation.
Then the ad expansion (\ref{(9)}) produces the next interaction $J^{(k+1)}$, which is again given by a sum of graphs 
$\sum_g J^{(k+1)}(g)$. This is a recursive construction. As in the first step, we may use the condition of nonvanishing commutators to enforce connectivity. Unwrapping
the expansions, one finds that each $g$ corresponds to a sequence of spin flips at a set of sites that is nearest-neighbor connected. 
At each stage of the procedure, quasilocality of the interaction is preserved.

The set of resonant blocks in the $k^{\text{th}}$ step is obtained by taking the connected components of the set of sites that belong to resonant graphs $g$.
In order to preserve the diluteness of the resonant region (an essential part of the notion of quasilocality),
one needs to maintain uniform exponential decay on the probability that $g$ is resonant. 
To this end, one proves that for $s = \tfrac{2}{7}$, there is a bound
\begin{equation}\label{(11)}
\mathbb{E}\,|A^{(k)}_{\sigma\tilde{\sigma}}(g)|^s \le\gamma^{s|g|} \mathbb{E} \prod_{\tau \tilde{\tau}\in g} \left|E^{(j)}_{\tau} - E^{(j)}_{\tilde{\tau}}\right|^{-s} \le (c\gamma)^{s|g|}
\end{equation}
on the fractional moment of the rotation generator.
Then Markov's inequality implies that
\begin{equation}\label{(12)}
P\left(|A^{(k)}_{\sigma\tilde{\sigma}}(g)| > (\gamma/\varepsilon)^{|g|}\right)
\le (c\varepsilon)^{s|g|}.
\end{equation}
The number of graphs containing a given site is exponential in $|g|$, so (\ref{(12)}) controls the sum over collections of graphs connecting one site to another. Then it is clear that resonant blocks are dilute and do not percolate. As in the first step, rotations are performed in resonant blocks to diagonalize the Hamiltonian there.

\subsubsection{Going beyond the forward approximation}
However, there are complications in obtaining (\ref{(11)}) if the graph $g$ involves a significant number of repeated spin flips. With repeated spin flips, energy denominators can appear to a high power, or there can be a large number of relations between them. If this is the case, then the fractional moment will no longer be finite, because of the lack of integrability of $|h|^{-sp}$ for $p\ge1/s$. In effect, (\ref{(11)}) is valid only in the forward approximation, wherein graphs looping back to previously visited sites are not allowed. In order to handle the troublesome graphs, one relies on previously obtained bounds $|A^{(j)}_{\sigma\tilde{\sigma}}(\tilde{g})| \le (\gamma/\varepsilon)^{|\tilde{g}|}$ for $j<k$. By induction, these can be used in the nonresonant region, and they lead to exponentially decaying estimates on 
$A^{(k)}_{\sigma\tilde{\sigma}}(g)$. Such estimates would tend to degenerate with $k$, were it not for the fact that graphs with many repeated spin flips do not span as great a distance as ``self-avoiding'' ones. One can, however, obtain uniform decay in the span of the graph, and this is sufficient for quasilocality and convergence in subsequent steps.

Similar arguments allow control over the factorial number of terms in the ad expansion (\ref{(9)}). Even in the first step, there are $\sim n!$ choices for indices
$\{i_1,\ldots,i_n\}$ such that $\prod_{p=1}^n(\text{ad}\,A(i_p))J(i_0) \ne 0$.
This is compensated in perturbation theory by the inverse factorials in (\ref{(9)}).
However, in the Markov inequality, one needs some improvement, because the available $1/n!$ has to control two sums: (1) the sum over graphs in the ad expansion and (2) the sum over events 
$\{A^{(k)}_{\sigma\tilde{\sigma}}(g) \text{ is resonant}\}$. 
Improvement is possible for graphs with few repeated spin flips, because in that case the number of graphs grows only as an exponential in $n$.
As explained above, inductive bounds are sufficient when working with graphs with many repeated spin flips.

\subsubsection{The role of dimensionality}
\begin{figure}[h]
\centering
\includegraphics{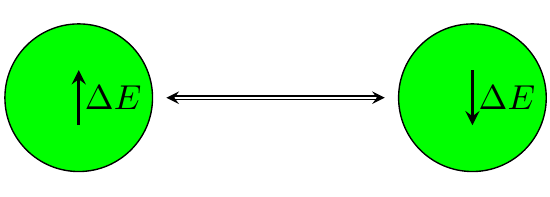}
\caption{A block-block resonance.\label{figblock}}
\end{figure}

In later steps, graphs may connect resonant blocks with nearby sites or with other blocks. 
For graphs connecting different blocks, the fractional-moment bound depends on having some control over the probability that an energy difference in a block is close to that of a given nearby transition. 
See Fig. \ref{figblock}.
One can obtain the necessary bounds using the level-spacing assumption (\ref{(2)}). See \cite{imbrie2014many} for details.
For a block $B$ of diameter $L$, connections to sites at a distance $d$ have size $\sim \gamma^d$. This needs to be smaller than $2^{-L} = 1/( \text{number of states in }B)$ if they are to be treated perturbatively. So $B$ needs to be expanded to $\bar{B}$, which includes all sites within a distance $d =  O(L)$  of $B$. Then exact diagonalization in $\bar{B}$ will eliminate the problematic nearby interactions. Expanding the blocks leads to enhanced connectivity, as nearby blocks have to be combined. The extended blocks may be thought of as connected clusters of a percolation problem. One can show that the probability that sites $i$ and $j$ lie in the same extended block decays as $(\gamma^{\kappa})^{1+(\log|i-j|)^2}$. This provides quantitative control over the size of regions where rotations far from the identity may be required, \textit{i.e.}
a quantitative statement of quasilocality.

In dimension 2 or more, the proof of MBL breaks down, because for $L$ large, there is no $d$ such that $\gamma^d$ is smaller than $2^{-\text{volume}(\bar{B})}$. It is argued in \cite{Roeck2016} that this issue destabilizes the MBL phase at very long times in dimension 2 or more. We postpone further comments about this problem to Sec. \ref{sec:Discussion}.

\subsubsection{Eigenstate Labels and Local Integrals of Motion}
Taking the limit as $k$ tends to infinity, all off-diagonal entries of the Hamiltonian are eliminated. 
Thus the cumulative rotation from the procedure diagonalizes $H$. 
The columns of this rotation matrix are the many-body eigenfunctions.

Let us discuss how diagonalization of the Hamiltonian with quasilocal rotations leads to a complete set of local integrals of motion. As explained above, diagonalizing $H$ can be accomplished with small, quasilocal rotation generators throughout most of $\Lambda$. One may call this region the perturbative region. 
There is a dilute, nonpercolating set of resonant blocks where rotations far from the identity may be required. The resonant blocks constitute the nonperturbative region. In the perturbative region, the cumulative rotation is close to the identity. 
Thus each eigenstate arises as a small rotation performed on one of the tensor product basis vectors; hence one can unambiguously label it by the spin configuration $\sigma$ that defines that basis vector. 
Adding in the nonperturbative region, it is evident that the eigenstate labels have to be supplemented with additional information in each resonant block. 
The need for state labels arises when rotations are performed in blocks to diagonalize the Hamiltonian there. 
The choice of labels is rather arbitrary, as we saw earlier in the $2 \times 2$ example (\ref{(1.1)}). In \cite{imbrie2014many}, the labels are called ``metaspins.'' 
For a block of size $n$, the metaspin takes $2^n$ values. The following discussion assumes a choice of 1:1 correspondence between the metaspins of a block and the spin configuration $\sigma$ in that block.

Having assigned spin labels to the eigenstates of $H$, one may proceed to define the local integrals of motion. 
Let $\Omega$ be the rotation that diagonalizes $H$, so that $\tilde{H} = \Omega^* H\Omega$ is diagonal. Note that the columns of $\Omega$ are the eigenvectors. 
As the eigenvectors are labeled by spin configurations $\sigma$, one may use $\sigma$ as the column index for $\tilde{H}$. 
Similarly, the rows of $\tilde{H}$ are indexed by spin configurations. Thus $\tilde{H}$ is a diagonal matrix that can act on vectors in the original tensor product basis. 
Although there is some arbitrariness in how this action is defined, the ambiguity is limited to a dilute set of resonant blocks.
Note that the spin operators $S^z_i$ are also diagonal in this basis. Therefore, $[\tilde{H},S^z_i] = 0$ for each $i\in \Lambda$.
Transferring the rotations in this equation to the spin operators, one obtains that $[H,I_i]=0$, where
\begin{equation}\label{tau}
I_i \equiv \Omega S^z_i \Omega^*.
\end{equation}
The rotations that diagonalize $H$ have been used to rotate the spin operators $S^z_i$ into local integrals of motion $I_i$.

Note that $I_i$ is defined by acting on $S^z_i$ with the rotation opposite to the one used to diagonalize $H$. The forward rotation $\Omega^*S_i^z\Omega$ produces the matrix elements of $S^z_i$ between eigenstates. In particular, 
$[\Omega^*S_i^z\Omega]_{\sigma\sigma}$ gives the expectation of $S_i^z$ in the eigenstate with label $\sigma$. Away from resonant blocks, this rotation is close to the identity. Thus the eigenstate with label $\sigma$ resembles the basis vector with the same label: we have that $\langle S_i^z\rangle_\sigma = \sigma_i + O(\gamma^\kappa)$ for some $\kappa>0$. The probability that $i$ lies in a resonant block is $O(\gamma^\kappa)$ as well. This explains the conclusion (\ref{(1.4)}) of Theorem \ref{thm:1}.

One may get another perspective on this construction of local integrals of motion by considering the analogous problem in the context of the Anderson model \eqref{eq:Hamiltonian} on a rectangle $\Lambda \subset \mathbb{Z}^d$, with weak hopping $0<\gamma \ll 1$.  %consider the Hamiltonian
%\begin{equation}
%H_{\text{A}} = -\gamma \Delta + v,
%\label{anderson}
%\end{equation}
%where $\Delta$ is the lattice Laplacian, and $v$ is multiplication by a random potential $v_x$, $x\in \Lambda$. 
In \cite{Imbrie2016b}, a sequence of quasilocal rotations were constructed to diagonalize $H_{\text{A}}$. This work served as a model case on which the many-body construction of \cite{imbrie2014many} was based.
Paralleling the constructions in (\ref{(5)})-(\ref{(12)}), one may specify that a site $x$ is resonant if $|\epsilon_x-\epsilon_y| < \varepsilon = \gamma^{1/20}$ for any $y$ with $|x-y| = 1$. Then, away from the set of resonant sites, one may define an antisymmetric matrix
\begin{equation}\label{rot}
A_{xy} = \frac{J_{xy}}{\epsilon_x - \epsilon_y}, \text{ for } x \ne y,
\end{equation}
where 
\begin{equation}\label{J}
J_{xy} = \begin{cases}
\gamma, &\text{if } |x-y| = 1; \\
0, &\text{otherwise.}
\end{cases}
\end{equation}
contains the hopping terms of $H_{\text{A}}$.
Then $A$ may be used to generate the first-step rotation. On the dilute set of resonant blocks, large rotations may be needed to diagonalize the Hamiltonian locally. Continuing with the multiscale procedure outlined above in the many-body case, one obtains again a rotation matrix $\Omega$ generated via quasilocal operators. Away from resonant blocks, the rotation is close to the identity, so as before one can label eigenstates by the original basis vectors (allowing for some arbitrariness in the choice of 1:1 correspondence in resonant blocks). Writing  $\psi_x$ or $|\psi_x\rangle$ for the eigenfunction with label $x$, one finds that $|\psi_x(y)| \le \gamma^{|x-y|/2}$ for $|x-y| > R$. Here $R$ depends on the disorder, but $\text{Prob}(R>L) \le \varepsilon^{L}$.

As above, one finds that $\tilde{H}_{\text{A}} = \Omega^*H_{\text{A}}\Omega$ is diagonal, and $\Omega_{yx} = \psi_x(y)$.
Let $|x\rangle$ denote the basis vector which is 1 at $x$ and 0 elsewhere. Then $|x\rangle\langle x|$ is the projection onto functions supported at $x$. As it is a diagonal operator, it commutes with 
$\tilde{H}_{\text{A}}$. As in (\ref{tau}), one can define 
\begin{equation}\label{tautilde}
\tilde{I}_x \equiv \Omega |x\rangle\langle x| \Omega^* = 
|\psi_x\rangle\langle \psi_x|,
\end{equation}
and then $[H_\text{A},\tilde{I}_x] = 0$. So in this case the local integral of motion $\tilde{I}_x$ is simply the projection onto the subspace spanned by the eigenstate with label $x$.

In an analogous fashion, one may rewrite (\ref{tau}) in the many-body case. Write $|\sigma\rangle$ for the basis vector associated with the spin configuration $\sigma$, and  $|\psi_\sigma\rangle$ for the associated interacting eigenfunction. Then
\begin{equation}\label{S}
S_i^z = \sum_\sigma \text{sgn}(\sigma_i)|\sigma\rangle\langle\sigma|
\end{equation}
and
\begin{equation}\label{tautau}
I_i = \sum_\sigma \text{sgn}(\sigma_i)|\psi_\sigma\rangle\langle\psi_\sigma|.
\end{equation}

\section{Construction of conservation laws: numerical schemes}\label{sec:numericalConstruction}
Beside being the suitable framework in which to prove rigorously the occurrence of MBL, the schemes to diagonalize quantum Hamiltonians iteratively are amenable to be implemented numerically~\cite{monthus2016flow}. They give prescriptions to construct sequences of rotated Hamiltonian $H^{(n+1)}=U^\dag_n H^{(n)} U_n$ converging to a diagonal form; each such prescription defines a set of conserved quantities (as in \eqref{tau}), that are expected to be quasilocal in the MBL phase. Similar iterative schemes are also at the basis of the RG procedures used to characterize the MBL phase: conservation laws emerge naturally from this setting as well. Approximately conserved operators can also be obtained by means of variational procedures minimizing the (Frobenius) norm of their commutator with the Hamiltonian. We shortly review these numerical approaches in the following.

\subsection{Conserved quantities obtained from diagonalizing flows}
Unitary rotations can be repeatedly applied to diagonalize the Hamiltonian by means of the systematic elimination of its strongest off-diagonal term with respect to some chosen basis, in analogy with the Jacobi algorithm for the diagonalization of matrices~\cite{Jacobi1846}. The extension of the Jacobi algorithm to fermionic Hamiltonians has been proposed in~\cite{White2002} and discussed, with specific reference to the MBL problem, in~\cite{rademaker2015}. The off-diagonal terms of the Hamiltonian are in this case strings of fermionic operators (in the local basis of single-particle states) that cannot be rewritten as products of only number operators. Each partial rotation eliminating an off-diagonal string generates new strings of equal or longer length, allowing to eliminate the off-diagonal terms order by order in the length of the corresponding operator. The transformation preserves the many-body Hilbert space as no degree of freedom is integrated out after each partial rotation: as a consequence, terms eliminated at a given step are typically regenerated at subsequent steps, with renormalized coefficients whose distribution flows to zero if the method converges.

Alternative, continuous diagonalizing flows have been discussed in~\cite{Pekker2016FixedPoints,Pekker2016flow, monthus2016flow}. The Wegner-Wilson flow has been exploited in \cite{Pekker2016FixedPoints} to diagonalize the Heisenberg chain \eqref{eq:XXZ} with $J=J_z=1/4$ and random fields $h_i \in \quadre{-w/2,w/2}$. The distribution $\mathcalnew{F}_{r, w}(h)$ of the coefficients $h_{i j \cdots}$ in \eqref{eq:EffMod} is analyzed as a function of the maximal distance $r$ between the sites $i, j, \cdots$. Evidences of a ``flow'' of the distribution $\mathcalnew{F}_{r, w}(h)$ with $r$  towards a $1/f$ law are given for the MBL phase, while the distribution appears to be scale-free (independent on $r$) in the critical regime. 
An adaptation of the Toda flow to random spin chains, having the advantage of preserving the sparsity of the Hamiltonian, has been proposed in~\cite{monthus2016flow}.

\subsection{Conservation laws emerging from the RG schemes}
The RG approaches to MBL are a direct extension of the Strong Disorder Renormalization Group (SDRG) scheme developed to capture the low-temperature thermodynamical properties of random magnets~\cite{DasguptaMa1980,Fisher1992, MonthusIgloi}. In its conventional form, the renormalization is performed on the system's Hamiltonian by means of an iterative elimination of its largest coupling constant, whose magnitude $\Omega$ defines the decreasing energy-cutoff. The corresponding spin subsystem is diagonalized, projected onto the lowest local energy state (or manifold), and decimated from the chain by introducing an effective coupling between its neighboring spins (that accounts perturbatively for the transitions induced by quantum fluctuations involving the virtual occupation of the excited state of the subsystem). The outcome of the RG procedure is an effective ground state, built as an approximate tensor product of few-spins lowest-energy states.

This procedure has been adapted in~\cite{pekker2013hilbert,vosk2013dynamical} to access the properties of the exited states of a one dimensional interacting transverse field Ising model
\begin{equation}\label{eq:RandomIsing}
 H=\sum_{i}\tonde{ J_i^z S^z_i S^z_{i+1}+ h_i S^x_i +  J^x_i S^x_i S^x_{i+1}},
\end{equation}
with uniformly distributed couplings $J^z_i, h_i$ with comparable size, and perturbatively small interactions $J_i^x$. The generalization consists in keeping track of the projection of the decimated subsystem on both its lowest and higher energy manifold. A full set of approximate eigenstates is thus constructed by progressively resolving energy gaps of the order of the running cutoff $\Omega$ \footnote{This approach admits a natural dynamical interpretation~\cite{Vosk:2013kq,vosk2013dynamical}, as the renormalized Hamiltonian can be interpreted as the operator giving the effective dynamics at the largest time-scales, when the fastest degrees of freedom oscillating at the scale $t \sim \Omega^{-1}$ have been integrated out. The phenomenological Hamiltonian \eqref{eq:EffMod} is recovered as the fixed point of the renormalization.}.

Conserved pseudospin emerge naturally from the RG scheme, as the operators whose eigenvalues $\pm 1$ label the energy subspaces selected at each RG step. For example \cite{altman2015review}, assume that at a given RG step the strongest coupling in \eqref{eq:RandomIsing} is $|J^z_j|= \Omega$, and let $H_{\Omega}=  J^z_j S^z_j S^z_{j+1}, \; V= H-H_{\Omega}$, and $v$ be the magnitude of the largest coupling in $V, (v < \Omega)$. The local two-sites Hilbert space at $j, j+1$ is partitioned into two almost degenerate energy manifolds (spanned by $| \uparrow \uparrow \rangle, |\downarrow \downarrow \rangle $ and $|\uparrow \downarrow \rangle \pm |\downarrow \uparrow \rangle $, respectively) separated by a gap of order $\Omega$.
The decimation is performed by means of a unitary transformation $H' \equiv e^{i A} H e^{-i A}$ chosen in such a way that $\quadre{H', H_\Omega}=0$; 
following the standard SDRG, this program is realized perturbatively by expanding the generator $A$ to lowest order in $v/\Omega$. Two different renormalized Hamiltonians are obtained projecting the second order truncation of $H'$ into the subspaces corresponding to a fixed eigenvalue of $S^z_j S^z_{j+1}$, 
\begin{equation}\label{eq:IsingSpinEffective}
\begin{split}
 H^\pm_{eff} &= V'\pm \frac{1}{J_j^z} \tonde{h_{j+1} J^x_{j+1}S^x_{j+2} + h_{j} J^x_{j-1}S^x_{j-1}}\\
 &\pm \frac{1}{J_j^z}\tonde{ h_{j} J^x_{j+1} \tilde{S}_j^x S^x_{j+2} + h_{j+1} J^x_{j-1} S^x_{j-2}  \tilde{S}_j^x }\\
 &+\tonde{J^x_j \pm \frac{h_j h_{j+1}}{ J_j^z}} \tilde{S}_j^x \pm \frac{J^x_{j-1} J^x_{j+1}}{ J_j^z} S^x_{j-2}\tilde{S}_j^x S^x_{j+2},
 \end{split}
\end{equation}
where $V'$ contains all terms in $V$ not involving the spins at the sites $j, j+1$ and $\tilde{S}_j^x$ is an effective spin operator breaking the internal degeneracy in the subspaces. Analogous steps are performed when the largest coupling is a local field. As a consequence of the projection, the effective dynamics within each energy manifold decouples. The operator $\tilde{I}_j=e^{-i A} (S^z_j S^z_{j+1}) e^{i A}$ is an emergent conserved quantity for $H$ (approximate since its dynamics is suppressed with the decimation), and the full set of approximate many-body eigenstates is clustered into two subgroups labeled by its eigenvalues, and separated by a gap of order $\Omega$. At the end of the procedure, the full spectrum is constructed with the resulting eigenstates given as product states in the basis of the approximate conserved quantities. Since resonances between distant degrees of freedom are not accounted for, %\footnote{The reason for this is that the effective interactions between distant spins appears after a number of RG step that is at least as large as the distance between the spins; thus, the transition hybridizing distant resonant spins having comparable local couplings at some scale $\Omega$ are not accounted for  since the corresponding degrees of freedom are decimated at earlier steps}, 
the scheme is expected to provide accurate results in the strong disorder limit.

An alternative real space RG scheme has been proposed in~\cite{monthus2015integrals}, based on a block renormalization running on an increasing sequence of length scales $L_k= 2^k$. At each step $k$, a conserved quantity $\tau_i^{(k)}$ is produced in parallel for each of the partitioning block, as a pseudospin operator whose eigenvalues label the degenerate energy manifolds obtained diagonalizing an intra-block Hamiltonian. %\footnote{Consider the simplest example of a quantum Ising chain in transverse filed: at the first step, the chain is partitioned into blocks of two sites, and the intra-block Hamiltonian is chosen to be $H^{(1)}_i \equiv h_{2i-1} \sigma^x_{2i-1}+ J^z_{2i-1}  \sigma^z_{2i-1} \sigma^z_{2i}$, which commutes with $\sigma^z_{2i}$. The corresponding four eigenstates can be  labeled by two quantum numbers, associated respectively to a pseudo-spin $\tau^z_{2i-1}$ and to $\sigma^z_{2i}$.}. 
The interaction between the blocks is renormalized assuming that the pseudospin operators are constant of motion, producing an effective Hamiltonian at a larger length scale. 

\subsection{Local operators that are approximately conserved}
 If quasilocal conserved operators exist, a good approximation for them can be obtained by cutting off the exponentially decaying tails, truncating the operators to a finite region of size $L \gtrsim \xi_{op}$. The resulting operators are \emph{approximately} conserved local operators, or ``ALIOMs'', as they commute with the Hamiltonian up to terms that are exponentially small in $L$. In \cite{Abanin2016Explicit}, a set of ALIOMs supported on a finite interval of length $L$ is constructed by expanding them in a complete basis of local operators, and by fixing the coefficients by direct minimization of the Frobenius norm of the commutator with the Hamiltonian. The problem is equivalent to an eigenvalue problem for a $(4^L \times 4^L-1)$ matrix, whose low-eigenvalue manifold maps to the set of approximately conserved operators and their products. Similar quantities have been constructed heuristically in \cite{Pollet2016}, where a quantum Monte Carlo approach to the highly excited MBL eigenstates is proposed. 
 
 An alternative approach has been adopted in \cite{Soumya2015SingleParticlePerspective}, by diagonalizing the one-body density matrix defined on individual many-body eigenstates. In the MBL phase, the resulting eigenvalues $\grafe{\tilde{n}_\alpha}_{\alpha=1, ..., L}$ have been found to be close to either $0$ or $1$, with eigenvectors that are spatially localized.
 A step-like discontinuity in the occupation spectrum is present in the MBL phase, analogous to the one of Fermi liquids. This suggests that the operators $\tilde{n}_\alpha= c^\dag_\alpha c_\alpha$, $\alpha$  eigenstates of the reduced density matrix have large overlap with the exactly conserved LIOMs.
 
 %the operators are obtained as a linear combination of a subset of the local operator basis (including only terms supported at most on nearest neighboring sites), and used in conjunction with quantum Monte Carlo techniques to access to the properties of highly excited states in the MBL phase.

\section{Discussion and conclusive remarks}\label{sec:Discussion}
In this section we discuss recent works on the extension of the idea of LIOMs beyond their original inception. Although some of these works are not universally accepted in the community, we discuss them anyway for their possible important implications.

\subsection{Many-body mobility edges and LIOMs} 
Firstly, we should mention that it is not completely clear how to reconcile the interplay of an extensive mobility edge (separating the lower-energy, localized many-body eigenstates from the higher energy, extended ones) with the existence of LIOMs. In the approximate perturbative constructions of \cite{Basko:2006hh,gornyi2005interacting,rms-IOM,laumann2014many,Baldwin2016qRem}, the temperature dependence of the critical interaction is obtained by considering, in the perturbative analysis, typical non-interacting states at the given temperature, which are treated within the forward approximation (or its analogues). This treatment leads to a temperature or energy-density dependent critical value of the interaction $\lambda_c(W,T)$ (where $W$ is the disorder), which, by parallel with the critical disorder in Anderson localization, defines a mobility edge $\epsilon(\lambda,W)$.

Signatures of mobility edges have been found in numerics of infinite-dimensional and 1d systems \cite{Kjall:2014fj,alet2015,laumann2014many}. Recent works \cite{de2016absence} maintain however that, for any $\lambda>\min_{T}\lambda_c(T)$, rare spontaneous local fluctuations in the energy-density within putative localized states (termed ``bubbles") are sufficient to restore the conductivity of the system at any temperature, as they can move resonantly through the system and act as a mobile bath. This argument rules out the scenario of a transition driven by temperature; it is reconciled with the perturbative treatment discussed in Sec.\ref{sec:PerturbationTheory} or analogous ones (which instead entail a finite-temperature transition) by noticing that bubbles are not captured within the forward approximation exploited in those contexts \footnote{Within the operator formalism of Sec. \ref{sec:PerturbationTheory}, transport driven by mobile bubbles would correspond to a divergence of the expansion \eqref{eq:Ansatz} due to a subsequence of operators having  support that is compact (with bounded index level $N$), but localized at increasingly far distance from the localization center of $I_\alpha$. These processes are not accounted for in the forward approximation.}. The authors of \cite{de2016absence} provided evidence for the absence of many-body mobility edges following from the ``bubbles" scenario, which is however still subject to debate.

If on the other hand a mobility edge is present, it is natural to expect that certain conservation laws are recovered once a projection onto the localized portion of the Hilbert space is performed. One step in this direction has been made in \cite{Nandkishore2016emergent} for a Heisenberg chain (although for values of the parameters corresponding to a fully MBL phase), by projecting the local spin operators onto a subspace $\mathcalnew{H}_{1-f}$ of the total Hilbert space $\mathcalnew{H}=\mathcalnew{H}_{1-f} \oplus \mathcalnew{H}_f$, spanned by a finite fraction $(1-f)$ of eigenstates.
Conserved quantities are obtained from the resulting projected operators by performing an infinite-time average of the operators evolved with the (non-local) Hamiltonian governing the dynamics in $\mathcalnew{H}_{1-f}$, in analogy with the recipe discussed in Sec.\ref{sec:TimeAv} of this review.
The resulting operators are argued to be local in the sense that their ``weight'' in compact regions of the chain remains finite in the thermodynamic limit, although there is a global dressing of the operators, whose total weight grows with $f$ \footnote{More precisely, the locality of the projected operators  ${{\rm O}}$ is understood in the following way: given the decomposition ${{\rm O}}= {{\rm O}}^A+{{\rm O}}^\perp$, where ${{\rm O}}^A$ is supported on some compact interval of the spin chain of size $N_A$, the ratio of the Frobenius norms $\lambda= \|{{\rm O}}^A\|^2 / \| {{\rm O}} \|^2 $ remains finite (of the order of $1-f$) as the thermodynamic limit is taken with $N_A$ kept fixed. However, $\| {{\rm O}}^\perp \|$ does not decay to zero exponentially as $N_A$ is increased, as it would be required for the operator to be quasilocal.}. 
%In a more recent work \cite{Nandkishore2016emergent}, conserved operators for an MBL Heisenberg chain in random fields are constructed by projecting out a subspace $\mathcalnew{H}_f$ of the Hilbert space spanned by a fraction $f$ of the eigenstates. The operators are conserved by the dynamics given by the (non-local) projected Hamiltonian, and are obtained by projecting onto $\mathcalnew{H}_f^\perp$ an extensive set of strictly local operators ${\rm O}_i$, and taking the infinite time average of the resulting operators with respect to the projected Hamiltonian. The resulting set of operators $\tilde{{\rm O}}_i$ is argued to be local in the following sense: given the decomposition $\tilde{{\rm O}}= \tilde{{\rm O}}^A+\tilde{{\rm O}}^\perp$ with $\tilde{{\rm O}}^A$ supported on some compact region of size $N_A$, the ratio of the Frobenius norms $\lambda= \|\tilde{{\rm O}}^A\|^2 / \| \tilde{{\rm O}} \|^2 $ remains finite as the thermodynamic limit is taken with $N_A$ kept fixed. The operators thus have a finite ''weight'' in compact regions of space; however, they are not quasilocal in the sense of Eq. \eqref{eq:QuasiLoc}, since $\| \tilde{{\rm O}}^\perp \|$ does not decay to zero exponentially as $N_A$ is increased. Rather, there is a ``global dressing'' whose weight scales as $f/N$ when $f$ is kept finite as the system size $N$ is increased, while it decays to zero when $f \to 0$ for $N \to \infty$. 
\subsection{Higher dimensions} Effects of the dimensionality on (even a fully) MBL region have been discussed in \cite{Basko:2006hh}. However, since in that work (as well as in \cite{rms-IOM}) only typical diagrams are considered, not much difference is seen in the MBL phase in different dimensions, nor in the critical properties (which in \cite{Basko:2006hh} are paralleled to those of a directed polymer). On the other hand, some non-typical effects can spoil the phenomenology of LIOMs: this was considered in a particular setup in \cite{Chandran2016beyond}, and in a more general setup in \cite{Roeck2016}. The starting point of both works is that of identifying (either by construction, or by estimating its probability of appearance) a rare region in which the disorder is anomalously low so that, if the region were typical, it would be ergodic. 

In \cite{Roeck2016}, rare ``ergodic grains'' within regions of localized spins, generated by atypical disorder fluctuations, were considered. Exploiting a random-matrix  description of such ``bath-like'' regions, it was argued that full MBL is destabilized by the ergodic grains in any dimension $d \geq 2$. This implies that transport is restored at large times, and that the non-perturbative construction of LIOMs discussed in Sec. \ref{sec:imbrie} would break down in dimension higher than one.

In the setup of \cite{Chandran2016beyond}, instead, such an ergodic region was the $d-1$ dimensional face of a cube of side $L$ of fully MBL spins (characterized by the existence of $L^d$ LIOMs). This region of size $L^{d-1}$ has typical level spacing $e^{-a L^{d-1}},$ which in $d>2$ is much smaller than the typical matrix element $e^{-b L}$ of an operator coupling a spin in the bulk with a spin on the boundary (irrespective of $a,b$). This implies that any spin in the bulk can thermalize using the boundary, and therefore the eigenstates should look thermal. However, eigenstates are the infinite-time limit of a finite-size system: it was argued in \cite{Chandran2016beyond} that a timescale diverging with $L$ arises, which hinders transport on macroscopic distances. This phenomenology is described, in the words of \cite{Chandran2016beyond}, by means of $l^*$-bits $I_\alpha^*$, namely approximately conserved, quasilocal operators, whose commutator with the Hamiltonian is not exactly zero but vanishing in the thermodynamic limit:
\begin{equation}
[I^*_\alpha,H]=O(e^{-bL}).
\end{equation}
This discussion can be framed in that on coupling MBL systems to an external bath. This is only tangential to the topic of this review, and we refer the interested reader to the existing literature \cite{nandkishore2014spectral, NandkGopala2016baths, SmallBath}.

\subsection{Conclusive remarks} 
The aim of this work was to illustrate how the notion of emergent integrability has clarified the phenomenology of MBL systems, and to review the main recipes proposed in the literature to construct the conserved operators. Besides the topics already mentioned in this section, there are other possibly relevant implications that have not been touched in the review. For instance, it may be interesting to re-consider the implications of MBL on well known protocols in quantum computation (such as the quantum adiabatic algorithm \cite{Altshuler:2010ct, Laumann2015}) in view of this characterization in terms of the emergent integrability. Moreover, although several constructive schemes are available, few attempts have been made so far to exploit the LIOMs to explicitly compute physically relevant quantities (see \cite{rm2016remanent} for an example). The existence of many non-trivial conserved quantities in quantum interacting, disordered systems has potential implications that are not only of theoretical interest but also of practical relevance, in view of the potential use of MBL systems in quantum devices; it is thus certainly worth to keep investigating the problem discussed in this review.

\begin{comment}
\begin{figure}
  \includegraphics[width=\columnwidth,<more options>]{<name>}%
  \caption{\label{<label name>}\col 
    <caption text>.}
\end{figure}

\begin{figure*}
%   \twocolcaption
%   \sidecaption
  \includegraphics[width=\textwidth]{}%
  \caption{\label{} 
  }
\end{figure*}

\begin{table}
  \begin{andptabular}[<table width>]{<column declaration>}{<caption>}%
    <table contents>\\
  \end{andptabular}
\end{table}
\begin{table}
  \begin{andptabbox}[<table width>]{<caption>}%
    <contents>
  \end{andptabbox}
\end{table}

\begin{table*}
  \begin{andptabular}[<table width>]{<column declaration>}{<caption>}%
    <table contents>\\
  \end{andptabular}
\end{table*}

\end{comment}

 \bibliographystyle{andp2012}
\bibliography{bibRef}

\end{document}